\documentclass[useAMS,usenatbib,referee]{biom2arxiv}

\usepackage{bbm}
\usepackage{bm}
\usepackage{amsmath, mathtools}
\usepackage{graphicx}
\usepackage{xr}
\usepackage{url}
\usepackage{enumitem}
\setlist{nolistsep,noitemsep,topsep=0cm,after=\vspace{0cm},before=\vspace{0cm}}

\usepackage{color}

\DeclareMathOperator*{\argmax}{argmax}

\title{Randomization inference with general interference and censoring}

\author{
Wen Wei Loh$^{1,*}$\email{WenWei.Loh@UGent.be},
Michael G.\ Hudgens$^{2,**}$\email{mhudgens@bios.unc.edu},
John D.\ Clemens$^{3}$,
Mohammad Ali$^{4}$, and
Michael E.\ Emch$^{5}$ \\
$^{1}$Department of Data Analysis, Ghent University, Gent, Belgium \\
$^{2}$Department of Biostatistics, University of North Carolina, Chapel Hill, North Carolina, U.S.A. \\
$^{3}$Department of Epidemiology, University of California, Los Angeles, California, U.S.A. \\
$^{4}$Department of International Health, Johns Hopkins University, Baltimore, Maryland, U.S.A. \\
$^{5}$Department of Geography, University of North Carolina, Chapel Hill, North Carolina, U.S.A.
}

\begin{document}

\label{firstpage}

\begin{abstract}
Interference occurs between individuals when the treatment (or exposure) of one individual affects the outcome of another individual. Previous work on causal inference methods in the presence of interference has focused on the setting where a priori it is assumed there is `partial interference,' in the sense that individuals can be partitioned into groups wherein there is no interference between individuals in different groups. 
\citet*{Bowers:2012aa} and \citet*{Bowers:2016aa} 
consider randomization-based inferential methods that allow for more general interference structures in the context of randomized experiments. 
In this paper, extensions of Bowers et al. which allow for failure time outcomes subject to right censoring are proposed. 
Permitting right censored outcomes is challenging because standard randomization-based tests of the null hypothesis of no treatment effect assume that whether an individual is censored does not depend on treatment. 
The proposed extension of Bowers et al. to allow for censoring entails adapting the method of 
\citet*{Wang:2010aa} for two sample survival comparisons in the presence of unequal censoring. 
The methods are examined via simulation studies and utilized to
assess the effects of cholera vaccination in an individually-randomized trial of $73,000$ children and women in Matlab, Bangladesh.
\end{abstract}


\maketitle
\section{Introduction}

Interference arises when an individual's potential outcomes depend on the treatment status of others.
Assuming interference is absent when 
assessing the causal effect of a treatment on an outcome 
may be scientifically implausible in certain settings. 
For example, in the study of infectious diseases, 
whether one individual receives a vaccine may affect whether another individual becomes infected or develops the disease.
Motivated by infectious diseases and other settings where individuals interact, many existing causal inference methods have been extended to allow for interference; see \citet{Halloran:2016aa} for a recent review.  

Some previous work on causal inference methods in the presence of interference has 
assumed a priori that there is {\em partial interference} \citep{Sobel:2006aa},
that is, individuals can be partitioned into groups wherein there is no interference between individuals in different groups.
In this paper we consider the more general setting where interference between any two individuals may be assumed.
Recent approaches that allow for the presence of {\em general interference} when evaluating treatment effects include 
\citet*{Bowers:2012aa},
\citet*{Bowers:2016aa},
\citet{sussman2017elements} and
\citet*{Athey:2016aa}
among others. 
In randomized experiments where the treatment assignment mechanism is known, 
\citet*{Bowers:2012aa} (henceforth BFP) described 
how to carry out randomization-based (i.e., permutation or design-based) 
inference on parameters in {\em causal models}
which allow for general interference.
For an assumed causal model, 
a randomization-based approach 
entails constructing confidence sets for the causal parameters 
by inverting a set of hypothesis tests. 
An appealing aspect of randomization-based inference 
\citep[Chapter 2]{Rosenbaum:2002}
is that no assumption of random sampling from some 
hypothetical superpopulation is invoked.
Another benefit is the resulting $100(1-\alpha)\%$ confidence sets are exact, i.e., the probability the true causal parameters are contained in a confidence set
is at least the nominal level $1-\alpha$.
Moreover, in settings where possible interference is a priori assumed 
to have a specified network structure,
it is unreasonable to assume individual outcomes are independent,
such that standard frequentist approaches are not justified; 
in contrast randomization-based methods that allow for possible general interference readily apply.

In this article, we propose extensions of \citeauthor{Bowers:2012aa} to the setting where the response of interest is a failure time, and only the censoring time is observed for a subset of individuals due to right censoring.
In general, when there is right censoring randomization-based inference on the failure times is exact only when treatment does not affect the censoring times. 
The proposal to permit right censored observations thus
entails adapting the method of \citet*{Wang:2010aa} for two sample survival comparisons in the presence of unequal censoring. 
The remainder of this article is as follows.
In Section 2 notation is introduced, causal models are defined, and the randomization inferential procedure by Bowers et al. when there is no censoring is reviewed.
In Section 3 the proposed extension allowing for right censored outcomes is presented, and simulation study results are shown demonstrating the method approximately preserves the nominal size over a range of settings.
In Section 4 the methods are utilized to
assess the effects of cholera vaccination in an individually-randomized trial 
of $n=73,000$ women and children in Matlab, Bangladesh.
A brief discussion is provided in Section 5.

\vspace{-.25in}

\section{General Interference and Causal Models \label{sect:review}}

\subsection{General Interference}

Consider a finite population of $n$ individuals randomly assigned to either treatment or control.
For each individual $i= 1, \ldots, n$, let
$Z_i = 1$ if individual $i$ is assigned treatment and $Z_i = 0$ otherwise.
The vector comprising all treatment assignments 
is denoted ${\bf Z}= \left(Z_1, \ldots, Z_n \right)$.
The uppercase $\bf Z$ denotes the random variable corresponding to treatment assignment and the lowercase $\bf z$ denotes possible realizations of $\bf Z$. Let $y_i({\bf z})$ denote the potential outcome for individual $i$ that would be observed for treatment assignment $\bf z$; the observed outcome is denoted by $Y_i = y_i({\bf Z})$.
Let ${\bf y}({\bf z}) \!=\! \left(y_1({\bf z}), \ldots, y_n({\bf z}) \right)$ denote the vector of potential outcomes. The potential outcomes 
${\bf y}({\bf z})$ and ${\bf z} \in \{0,1\}^n$ are considered 
fixed features of the finite population of $n$ individuals.

Define the $n \times n$ {\em interference} matrix 
${\mathbbm A}$ with $(i,j)$ entry $A_{ij}$ for $i,j \in \{ 1,\ldots,n\}$ as follows.
Let $A_{ij} = 0$ for $i=j$. For $i \neq j$ let $A_{ij}=0$ if 
it is assumed a priori individual $j$ does not interfere with individual $i$;
otherwise let $A_{ij}=1$.
Note that $A_{ij}=0$ implies it is assumed a priori $y_i({\bf z})$ does not depend on $z_j$,
whereas $A_{ij}=1$ merely indicates the possibility that individual $j$ may interfere with individual $i$,
and does not necessarily imply $y_i({\bf z})$ depends on $z_j$.
Indeed, one of our primary inferential goals is to determine
whether such possible interference is present.
The definition of ${\mathbbm A}$ encodes the assumption that 
any spillover effects on individual $i$
may emanate only from individuals $j$ where $A_{ij}=1$, 
and not from those where $A_{ij} = 0$.
The exact relationship between $y_i({\bf z})$ and $\bf z$
is specified using a causal model described in the next section.
Let the {\em interference set} (i.e., neighbors) for individual $i$ 
be the set of individuals $j \in \{1,\ldots,n\} \setminus i$ where $A_{ij} =1$.
Denote the $i$-th row of ${\mathbbm A}$ by the vector ${\bf A}_i$,
and the size of the interference set
by the scalar $A_i = \sum_{j=1}^{n} A_{ij}$. %
Under partial interference, individuals can be partitioned into groups or clusters 
wherein there is no interference between groups, in which case ${\mathbbm A}$ 
can be expressed as a block-diagonal matrix with each block corresponding to a group. 
Under general interference, 
each individual is allowed to have their own possibly unique interference set,
so that there is no restriction on the structure of ${\mathbbm A}$.
Here and throughout ${\mathbbm A}$ is assumed known and invariant to treatment.

\subsection{Causal Models}

A (counterfactual) causal model
expresses the potential outcomes ${ y}_i({\bf z})$
as a parametric deterministic function of any treatment $\bf z$.
Following \citeauthor{Bowers:2012aa},
we consider a class of causal models which entail the composition of two functions. 
In particular, assume 
${ y}_i({\bf z}) = h\{{ y}_i({\bf 0}) | {\mathcal F}({\bf z}; \theta, {\mathbbm A})\}$
for user-specified functions $h$ and ${\mathcal F}$,
with ${ y}_i({\bf 0})$ denoting the potential outcome under the {\em uniformity trial} \citep{Rosenbaum:2007aa} where no one receives treatment.
The function ${\mathcal F}({\bf z}; \theta, {\mathbbm A})$ 
takes as its arguments
the treatment vector ${\bf z}$, 
causal parameter $\theta$, and 
interference matrix ${\mathbbm A}$.
The dependence of ${\mathcal F}({\bf z}; \theta, {\mathbbm A})$ on $i$ is left implicit notationally as it is implied under the specified causal model. For notational simplicity we write ${\mathcal F}={\mathcal F}({\bf z}; \theta)$, with the dependence on ${\mathbbm A}$ implicit. The specification of $\mathcal F$ determines how an individual's potential outcomes differ
across different treatments ${\bf z}$ and different values of the parameter $\theta$,
and includes, but is not limited to, 
how direct and spillover effects propagate.
The link function $h$ is a one-to-one function mapping
${ y}_i({\bf 0})$ to $y_i({\bf z})$ for a specified ${\mathcal F}$; in particular,
the uniformity trial potential outcomes can be determined from the observed data under a specified causal model by
${ y}_i({\bf 0}) = h^{-1} \left\{ { Y}_i | {\mathcal F}({\bf Z}; \theta) \right\}$,
where $h^{-1}$ is the inverse of $h(a|b)$.

In practice, prior beliefs or background knowledge may be used to inform the choice of ${\mathcal F}$ and $h$. We consider two specific causal models, defined in \eqref{eq:causalmodel:add} and \eqref{eq:causalmodel:BFP} below, and assume $h\!\left(a | b \right)=a\exp(b)$, although the proposed methods are general and apply to other forms of ${\mathcal F}$ and $h$.
Denote the number and proportion of individual $i$'s neighbors assigned to treatment by $T_i = {\bf A}_{i}{\bf Z}^T$ and $G_i = {T_i}/A_i$ respectively;
here $A_i=0$ implies $T_i=G_i=0$. Note that $T_i$ and $G_i$ depend on $\bf Z$, but this dependence is suppressed for notational convenience.
Let:
\begin{align}
{\mathcal F}_{{\rm add}}({\bf Z}; \delta,\tau) &=
\delta Z_i + \tau G_i; \label{eq:causalmodel:add}\\
{\mathcal F}_{{\rm BFP}}({\bf Z}; \delta,\tau) &= 
\delta + \log\left[ 1+(1-Z_i)\{\exp(-\delta)-1\}\exp(-\tau^2 T_i) \right].
\label{eq:causalmodel:BFP}
\end{align}
Under both causal models, 
the effect of treatment ${\bf Z}$ on the outcome for individual $i$ takes the form of 
a bivariate treatment:
$Z_i$ is the (individual) treatment received, and
$G_i$ (or $T_i$) is the proportion (or number) of individuals in the interference set treated.
The parameters $\delta$ and $\tau$ measure the extent to which
the potential outcomes increase or decrease, relative to $y_i({\bf 0})$, due to 
$Z_i$ and $G_i$ (or $T_i$).
Causal model \eqref{eq:causalmodel:BFP} was proposed by BFP and restricts interference to those who did not receive treatment, with the direct (or individual) effect parametrized to be larger in magnitude 
than the spillover (or peer) effect.
As both $G_i$ and $T_i$ depend only on the 
total number in the interference set treated, 
a peer effect homogeneity assumption is implied by these two causal models;
\citet{Hudgens:2008aa} refer to 
the assumption as {\em stratified interference}.
Causal models allowing for interference that does not occur via the summary $T_i$ can also be utilized within this framework. For example, we might posit ${\mathcal F} = \delta Z_i + \tau Z_{M_i}$ where $M_i = \argmax_{j:A_{ij}=1} A_j$ denotes the neighbor of individual $i$ having the biggest interference set. See \citet{ogburn2017causal} and \citet{sussman2017elements} for other causal models that allow for interference.
The next section describes how to carry out randomization inference
for the parameter $\theta=(\delta,\tau)$
under a specified $\mathcal F$.

\subsection{Randomization inference \label{sect:teststats_nocensoring}}

For a specified causal model $\mathcal F$, the uniformity trial potential outcomes under a null hypothesis $H_0: \theta=\theta_{0}$ can be determined from the observed data by ${ y}_i({\bf 0}) = { Y}_i \exp\left\{-{\mathcal F}({\bf Z}; \theta_0)\right\}$. 
In a randomized experiment where individuals are assigned treatment with equal probability, 
the uniformity trial outcomes should be similarly distributed 
between treatment ($Z=1$) and control ($Z=0$) groups \citep{Rosenbaum:2002} if $H_0$ is true and $\mathcal F$ is correctly specified.
Therefore the null hypothesis $H_0$ can be tested using a test statistic 
${\cal TS}({\bf Z}; \theta_{0})$
that compares the uniformity outcomes between treated and untreated individuals. 
For example, BFP used the two-sample Kolmogorov-Smirnov (KS) test statistic 
to compare the empirical distributions of the uniformity outcomes in the treatment and control groups.
\citet*{Bowers:2016aa} proposed a multiple linear regression model 
of the uniformity outcomes on $Z$ and $T$, 
using the resulting sum of squares of residuals as a test statistic.

For a chosen test statistic ${\cal TS}({\bf Z}; \theta_{0})$,
the plausibility of $H_0$ can be assessed by evaluating the frequency
of obtaining a value at least as `extreme' (from $H_0$) as the observed value,
over hypothetical re-assignments of $\bf Z$ under $H_0$.
Here and throughout a completely randomized experiment is assumed,
where the number assigned to treatment, denoted by 
$m=\sum_{i=1}^{n} { Z}_i$, is fixed by design.
The sample space of all hypothetical re-assignments $\bf Z$ is 
the set of vectors of length $n$ containing $m$ $1$'s and $n-m$ $0$'s, 
and is denoted by
$
\Omega =\left\{ 
	    {\bf z}: z_{i} \in \{0,1\}, i=1, \ldots, n,
	    \sum_{i=1}^{n} z_i = m
	    \right\}.
$
Each re-assignment occurs with probability $|\Omega|^{-1}$, 
so that a two-sided p-value may be defined as
$	{\rm pv}(\theta_{0}) \!=\! |\Omega|^{\!-1}
        \underset{{\bf z} \in \Omega}{\sum}
        {\rm I}\left\{{\cal TS}({\bf z};\!\theta_{0}) \!\geq\!
        {\cal TS}({\bf Z};\!\theta_{0}) \right\}$, 
where without loss of generality it is assumed the larger values of ${\cal TS}({\bf Z}; \theta_{0})$ suggest stronger evidence against $H_0$,
and ${\rm I}\{B\}=1$ if $B$ is true and $0$ otherwise.
When it is not computationally feasible to enumerate $\Omega$ exactly,
an approximation of $\Omega$ based on $\cal C$ random draws of $\bf z$ from $\Omega$ may be used to yield an approximate p-value, denoted by ${\rm pv}^{\cal C}(\theta_{0})$.

Confidence sets can be constructed by test inversion. The subset of $\theta_{0}$ values where ${\rm pv}(\theta_{0})$, or ${\rm pv}^{\cal C}(\theta_{0})$, is greater than or equal to $\alpha$ forms a $100(1\!-\!\alpha)\%$ exact confidence set for $\theta$. 
Confidence sets for individual parameters in $\theta$ can be obtained readily from a confidence set for $\theta$. For example, a $100(1\!-\!\alpha)\%$ confidence set for $\delta$ is given by all values of $\delta_0$ such that there exists some value of $\tau_0$ where ($\delta_0,\tau_0$) is in the $100(1\!-\!\alpha)\%$ confidence set for $(\delta,\tau)$.

It is important to note that each hypothesis test assesses the compatibility of the observed data with the assumed causal model $\mathcal F$ and assumed parameter values $\theta_0$ specified by $\mathcal F$ under the null. Rejection of the hypothesis only indicates that either $\mathcal F$ or $\theta_0$ is implausible. In some circumstances all feasible parameter values for an assumed causal model may be rejected, leading to an empty confidence set. This indicates all possible parameter values are implausible, implying that the assumed causal model provides a poor fit to the data.

\section{Right censored failure time outcomes \label{sect:rightcensored}}

Now suppose each individual's outcome is a (positive) failure time,
subject to right censoring if the individual is not followed long enough for failure to be observed.
For $i= 1, \ldots, n$, let $\tilde Y_i$ and $C_i$ denote the failure time and the censoring time respectively.
The failure time $\tilde Y_i$ is observed only if $\tilde Y_i \leq C_i$, so that the observed data are 
$Y_i = \min \{ \tilde Y_i, C_i \}$ and the failure indicator ${ D}_i = {\rm I}\{\tilde Y_i \leq C_i\}$.
The outcomes being right censored causes two complications for the randomization inference approach described in Section~\ref{sect:review}. First, the test statistic employed needs to account for right censoring; some possible statistics are discussed in Section~\ref{sect:teststats_yescensoring}. Second, the null hypothesis $H_0: \theta=\theta_{0}$ for a specified causal model ${\mathcal F}$ is no longer sharp in the sense that not all uniformity trial potential outcomes can be determined from the observed data under $H_0$. To see this, define ${ y}_i({\bf 0}) = { Y}_i \exp\left\{ - {\mathcal F}({\bf Z}; \theta_0) \right\}$, which can be determined from the observed data as in the previous section. Let $\tilde { y}_i({\bf 0}) = {\tilde Y}_i \exp\left\{ - {\mathcal F}({\bf Z}; \theta_0) \right\}$ denote the uniformity trial potential failure time for individual $i$ under $H_0$. 
For individuals who are not censored, $Y_i=\tilde Y_i$ implies $y_i({\bf 0}) = {\tilde y}_i({\bf 0})$, i.e., the uniformity trial potential failure time can be determined exactly under $H_0$ if ${ D}_i=1$. But for individuals who are censored, $\tilde Y_i$ is unobserved so that ${\tilde y}_i({\bf 0})$ is unknown under $H_0$. Nonetheless, it is known for these individuals that $Y_i<\tilde Y_i$; multiplying both sides of this inequality by $\exp\left\{ - {\mathcal F}({\bf Z}; \theta_0) \right\}$, it follows that $y_i({\bf 0}) < {\tilde y}_i({\bf 0})$. Thus the observed censoring times provide some information about the unknown failure times ${\tilde y}_i({\bf 0})$ for right censored individuals. In particular, $y_i({\bf 0})$ serves as a lower bound for $\tilde y_i({\bf 0})$ under $H_0$.
Because the null hypothesis is no longer sharp, the randomization testing approach in Section~\ref{sect:teststats_nocensoring} in the absence of censoring requires modification; the proposed approach is described in Section~\ref{sect:IPZ}. 

\subsection{Test statistics that accommodate right censoring \label{sect:teststats_yescensoring}}
The test statistics considered in Section~\ref{sect:teststats_nocensoring}
require modification to accommodate right censoring.
Instead of the KS statistic, the log-rank (LogR) statistic may be used to 
compare the right censored uniformity failure times in the treatment and control groups.
An analog of the multiple linear regression model is the
parametric accelerated failure time (AFT) model where
the log-transformed failure times are linear functions of the predictors.
In the following we consider a log-normal AFT model of the uniformity failure times 
given by
$
	\log \tilde y_i({\bf 0}) 
	= {\bf q}_i{\bm \beta} + \sigma \epsilon_i,
$
where 
${\bf q}_i \!=\! \left( 1, Z_i, G_i, Z_i G_i, A_i \right)$ and the errors $\epsilon_i$ are independent and normally distributed with mean zero and variance one.
(For the ${\mathcal F}_{\rm BFP}$ causal model, $G_i$ may be replaced by $T_i$.)
Following the likelihood ratio principle for testing,
a likelihood ratio permutation test is 
expected to be the most powerful test against certain alternatives
(see \citet{lehmann2005testing}, Chapter 5.9 for 
an example in the setting where there is no interference and no censoring).
Let ${\bf D} \!=\! \left( { D}_1, \ldots, { D}_n \right)$ denote the vector of failure indicators, 
and denote the log-likelihood by:
\begin{equation}\label{eq:aftloglik}
l({\bf Z}, {\bf D}; {\bm \beta}, \sigma, \theta_0) = \sum_{i=1}^{n} 
\left[ { D}_i \log \left\{ \phi(\epsilon_i) / (\sigma \tilde { y}_i({\bf 0})) \right\} 
+ (1-{ D}_i) \log \left\{ 1 - \Phi(\epsilon_i) \right\}
\right],
\end{equation}
where $\epsilon_i = \left\{ \log \tilde y_i({\bf 0}) - {\bf q}_i{\bm \beta} \right\}/\sigma$, and
$\phi$ and $\Phi$ are the standard normal density and distribution functions respectively;
see for example, Equation (6.25) of \citet{Collett:2003aa}.
Let $\hat {\bm \beta}$ and $\hat \sigma$ denote the maximum likelihood estimates (MLEs), and let $\tilde {\bm \beta}$ and $\tilde \sigma$ denote the MLEs for the `intercept-only' model, i.e., under the restriction ${\bm \beta}=(1,0,0,0,0)^T$. Then the log-likelihood difference is
${\rm LRaft} ({\bf Z}, {\bf D}; \theta_0) = 
l({\bf Z}, {\bf D}; \hat {\bm \beta}, \hat \sigma, \theta_0)
-
l({\bf Z}, {\bf D}; \tilde {\bm \beta}, \tilde \sigma, \theta_0)$.
In practice, $l({\bf Z}, {\bf D}; \hat {\bm \beta}, \hat \sigma, \theta_0)$ %
can be used in place of ${\rm LRaft} ({\bf Z}, {\bf D}; \theta_0)$
since $l({\bf Z}, {\bf D}; \tilde {\bm \beta}, \tilde \sigma, \theta_0)$ %
is constant with respect to $\bf Z$ for a fixed value $\theta_0$.
Note the AFT model should only be considered as a `working model,' used solely to generate a test statistic for a hypothesis testing procedure. 
Under the randomization-based framework, valid inference does not rely on this working model being correctly specified.
Rather, $l({\bf Z}, {\bf D}; \hat {\bm \beta}, \hat \sigma, \theta_0)$ %
can simply be viewed as a mathematical (scalar) summary of $\{{\bf y}({\bf 0}),{\bf D},{\bf Z}\}$ that is compared against other treatment assignments for assessing the plausibility of $H_0: \theta=\theta_0$.

\subsection{Correcting for right censored uniformity trial failure times \label{sect:IPZ}}
The randomization-based inferential procedures described in 
Section~\ref{sect:teststats_nocensoring}
do not necessarily yield tests that preserve the nominal size
in the presence of right censoring, 
even if the test statistics considered in 
Section~\ref{sect:teststats_yescensoring} 
are utilized. Randomization tests of no treatment effect on the failure times in the presence of censoring generally only preserve the nominal size when treatment does not affect the censoring times. 
To see this, consider for a moment the setting where there is no interference between individuals, 
so that each individual has two potential failure time outcomes $\tilde y_i(0)$ and $\tilde y_i(1)$,
and  two potential censoring times $c_i(0)$ and $c_i(1)$.  
Let $\tilde Y_i = \tilde y_i(Z_i)$ and $C_i = c_i(Z_i)$, and define $Y_i$ and $D_i$ as above.  
Consider testing the null hypothesis of no individual-level treatment effect, i.e., 
$H_0: \tilde y_i(0)=\tilde y_i(1)$ for $i=1, \ldots, n$,
using some test statistic which is a function of $\{ {\bf Y}, {\bf D}, {\bf Z} \}$ where ${\bf Y} = (Y_1, \ldots, Y_n)$ is the vector of observed outcomes. 
If we assume $c_i(0)=c_i(1)$, 
then under the null both $Y_i$ and $D_i$ will be the same regardless of treatment, 
allowing exact determination of the test statistic's sampling distribution 
by enumeration over all possible re-assignments in $\Omega$. 

However, when inverting a randomization test to construct a confidence set, null hypotheses corresponding to non-zero treatment effects on the failure times must also be tested. For such null hypotheses, the standard randomization testing approach described in Section~\ref{sect:teststats_nocensoring} cannot be used to determine a test statistic's sampling distribution under the null, because in general an individual's censoring indicator $D_i$ will not be fixed over all possible re-assignments ${\bf z} \in \Omega$, even if treatment has no effect on the censoring times. 
To see this, returning to the setting where there is interference consider the causal model ${\mathcal F}_{\rm add}$ and suppose $Z_i=1$ and $D_i=0$, i.e., individual $i$ is assigned treatment and is censored at time $Y_i$ with failure time $\tilde Y_i> Y_i$. 
Further assume treatment has no effect on the censoring times, so that the potential censoring time for individual $i$ equals $Y_i$ for all treatments ${\bf z} \in \Omega$. 
Now consider testing $H_0: \theta=\theta_0$ where $\delta_0>\log(\tilde Y_i/Y_i)$ and $\tau_0=0$. Then for treatment re-assignment ${\bf z}^\prime \in \Omega$ where $z_i^\prime=0$ it follows that $\tilde y_i({\bf z}^\prime) = \tilde y_i({\bf 0}) = \tilde Y_i \exp\left(-\delta_0\right) < Y_i$, i.e., individual $i$ would not be censored for treatment ${\bf z}^\prime$.
Thus, as will be demonstrated empirically in Section~\ref{sect:example_n256censoring_power} below, a randomization test that holds the set of censored individuals fixed over treatment re-assignments will not in general control the type I error. 
Instead, we propose the following randomization-based inferential procedure
that allows the set of censored individuals to vary over re-assignments. 

The procedure entails adapting the $\rm IP_Z$ permutation test by \citet{Wang:2010aa}. An outline of the procedure is as follows. First, ${\bf y}({\bf 0})$ is determined under $H_0$ using the specified causal model ${\mathcal F}$ and a test statistic from Section~\ref{sect:teststats_yescensoring} is evaluated at $\{{\bf y}({\bf 0}),{\bf D},{\bf Z}\}$. Second, the sampling distribution of the test statistic under $H_0$ over hypothetical treatment re-assignments is approximated by: (i) imputing the unknown uniformity trial failure times for censored individuals according to the assumed causal model $\mathcal F$ under $H_0$, and (ii) non-parametrically imputing censoring times using treatment group-specific Kaplan-Meier estimators of the censoring time distributions. No causal model is assumed for the censoring times.

The specific procedure is as follows. For a single observed dataset $\{ {\bf Y}, {\bf D}, {\bf Z} \}$, the following steps are carried out to test $H_0: (\delta, \tau) = (\delta_0, \tau_0)$:

\begin{enumerate}[label={\arabic*.},leftmargin=*]
	\item \label{step:adjustuniformity} 
	Determine the possibly right censored uniformity trial potential failure times under $H_0$,
	e.g., under the causal model ${\mathcal F}_{\rm add}$, $y_i({\bf 0}) = Y_i \exp\left\{-\left(\delta_0 Z_i + \tau_0 G_i\right)\right\}$. 
    Calculate the observed value of the chosen test statistic, e.g., the log-rank statistic, using
    $\{{\bf y}({\bf 0}), {\bf D}, {\bf Z} \}$.

    	\begin{enumerate}[leftmargin=*]
		\item Compute the Kaplan-Meier (KM) estimator of the distribution function of the uniformity failure times under $H_0$ using $\{{\bf y}({\bf 0}), {\bf D}\}$. 
		Denote the estimator by $\hat F_0(\cdot)$.

		\item For $z=0,1$, among individuals with treatment $Z_i=z$,
		compute the group-specific KM estimator of the censoring time distribution, 
		using the observed times $Y_i$ and censoring indicators $1-D_i$.
		Denote the estimators by $\hat S(\cdot|z)$.
		\end{enumerate}
		
    \item \label{step:draw_treatment}
    Randomly sample a new treatment assignment ${\bf z} \in \Omega$.
    
    \item \label{step:draw_failuretime}
    If $D_i=1$, set $\tilde y_i^\ast({\bf 0}) = y_i({\bf 0})$,
    where $y_i({\bf 0}) = \tilde y_i({\bf 0})$ is the observed uniformity failure time under $H_0$.
	    Otherwise if $D_i=0$, since $\tilde y_i({\bf 0})$ is unknown,
		sample a failure time from a truncated distribution with lower bound $y_i({\bf 0})$ as follows.
		Randomly draw $u \sim {\rm Uniform}[\hat F_0(y_i({\bf 0})),1]$.
		If $u \leq \hat F_0(\tilde y_{\max}({\bf 0}))$, 
		where $\tilde y_{\max}({\bf 0}) = \max_{i: D_i=1} \tilde y_{i}({\bf 0})$ is the maximum observed uniformity failure time,
		set the failure time as $\tilde y_i^\ast({\bf 0})=\hat F_0^{-1}(u)$;
		otherwise set $\tilde y_i^\ast({\bf 0})=\tilde y_{\max}({\bf 0})$.
    (The $\ast$ symbol distinguishes the imputed failure times from the unknown failure times $\tilde y_i({\bf 0})$ for those with $D_i=0$.)
	Determine the potential failure times under treatment $\bf z$ using the assumed causal model and the null parameter values, e.g., $\tilde y_i({\bf z}) = \tilde y_i^\ast({\bf 0}) \exp( \delta_0 z_i + \tau_0 g_i)$, where $g_i$ is the realization of $G_i$ under treatment ${\bf z}$.
		
    \item \label{step:draw_censoringtime}
    Sample a censoring time under treatment assignment $\bf z$, denoted by $c_i({\bf z})$, from $\hat S(\cdot|z_i)$ as follows.
    Randomly draw $v \sim {\rm Uniform}(0,1)$.
    Let $Y_{\max} = \max_{i: Z_i = z_i} Y_i$ be the maximum observed time among individuals with treatment $Z_i=z_i$.
    If $Y_{\max}$ is a censoring time, then the KM estimator of the censoring time distribution evaluated at $Y_{\max}$ is $\hat S(Y_{\max}|z_i) = 1$. Hence set the censoring time to be $c_i({\bf z}) = \hat S^{-1}(v|z_i)$. Otherwise if $Y_{\max}$ is a failure time so that $\hat S(Y_{\max}|z_i) < 1$, set the censoring time to be $c_i({\bf z}) = \hat S^{-1}(v|z_i)$ if $v \leq \hat S(Y_{\max}|z_i)$ and let $c_i({\bf z}) = Y_{\max}$ otherwise. Hence $c_i({\bf z}) \leq Y_{\max}$ so that any imputed potential failure time longer than $Y_{\max}$ will be censored at $Y_{\max}$.    
            
    \item \label{step:find_obsYD} 
    Determine the potential outcomes under treatment ${\bf z}$ as $y_i({\bf z}) = \min\{ \tilde y_i({\bf z}), c_i({\bf z}) \}$ and the failure indicators as ${ d}_i({\bf z}) = 1$ if $\tilde y_i({\bf z}) \leq c_i({\bf z})$ or $0$ otherwise.
    
    \item \label{step:find_samplingdist} 
    Determine the uniformity outcomes under treatment ${\bf z}$ using the same causal model as in step 1, e.g.,
    $y_i^\dag({\bf 0}) = y_i({\bf z})\exp\left\{-\left(\delta_0 z_i + \tau_0 g_i\right)\right\}$.
    (The $\dag$ symbol denotes the uniformity outcomes determined using $y_i({\bf z})$, which differ from the uniformity outcomes determined using $Y_i$ in step \ref{step:adjustuniformity})
    Compute the chosen test statistic using
    $\{{\bf y}^\dag({\bf 0}), {\bf d}({\bf z}), {\bf z} \}$,
    where ${\bf y}^\dag({\bf 0}) = \left( y_1^\dag({\bf 0}), \ldots, y_n^\dag({\bf 0}) \right)$
    and ${\bf d}({\bf z}) = \left( { d}_1({\bf z}), \ldots, { d}_n({\bf z}) \right)$ are vectors of length $n$.
    
    \item The sampling distribution of the chosen test statistic can be obtained by repeating steps 2 to \ref{step:find_samplingdist} The p-value for testing $H_0$ can be determined by comparing the resulting sampling distribution with the observed value of the chosen test statistic from step \ref{step:adjustuniformity}
        
\end{enumerate}

\subsection{Empirical evaluation of proposed tests \label{sect:example_n256censoring_power}}

In this section, the ability of the proposed procedure to better control the type I error in the presence of right censoring is assessed empirically. A simulation study is conducted as follows. 
The total number of individuals $n$ is set to 128, with exactly $m$ individuals assigned to treatment as in a completely randomized experiment.
For each individual $i$, the interference set ${\bf A}_{i}$ is generated once as follows:
(i) randomly draw the interference set size as $A_i\sim{\rm Poisson}(16)$;
(ii) sample without replacement $A_i$ values of $j \in \{1,\ldots,n\}\setminus i$ and set $A_{ij}=1$ for the sampled values of $j$; then (iii) set the remaining values of $A_{ij}$ to 0.
\begin{enumerate}[label=Step \arabic*.,leftmargin=*]\setcounter{enumi}{-1}
\item Sample the uniformity failure times as
$\log {\tilde y}_i({\bf 0})  \sim {\cal N}(\mu, \sigma^2)$, where $(\mu,\sigma^2)=(4.5,0.25^2)$.
\item Randomly draw an observed treatment assignment ${\bf Z}$ from $\Omega$.
Determine the failure time for individual $i$ with observed treatment ($Z_i,G_i$) by $\tilde Y_i \!=\! \tilde y_i({\bf 0}) \exp\left( \delta^{\dag} Z_i \!+\! \tau^{\dag} G_i \right)$
for $(\delta^{\dag},\tau^{\dag})=(0.7,2.8)$.
The values of $(\delta^{\dag},\tau^{\dag})$ are chosen so that
for $G_i>0.25$, 
the magnitude of the spillover effect is greater than the direct effect, i.e.,
$\tau^{\dag}G_i > \delta^{\dag}$.
The censoring times are then drawn from distributions that depend on treatment.
First the dropout times $\tilde C_i$ are randomly drawn from a lognormal distribution
$\log \tilde C_i \sim {\cal N}(\mu + \tau^{\dag} G_i, \omega^2)$, where $\omega^2=1-0.25^2$.
The administrative censoring time is defined as $C_i^{\prime}= \exp(\mu+2\sigma+\tau^{\dag})$.
If $Z_i=1$, set the censoring time to $C_i = \min \{ C_i^{\prime}, \tilde C_i\}$;
otherwise, assume there is no dropout and $C_i = kC_i^{\prime}$ for some specified proportion $k$.
Determine the observed outcomes $Y_i$ and failure indictors $D_i$ as defined above.
\item Under $H_0: (\delta_0, \tau_0)=(0.7,2.8)$, determine $y_i({\bf 0}) = Y_i \exp\left\{-\left( \delta_0 Z_i + \tau_0 G_i \right)\right\}$.
For the dataset $\{ {\bf y}({\bf 0}), {\bf D}, {\bf Z}\}$, 
carry out the LogR and LRaft tests, either holding $\bf D$ fixed over re-assignments, or using the proposed method in Section~\ref{sect:IPZ}.
The p-values ${\rm pv}^{\cal C}(\delta_{0},\tau_{0})$ are calculated with ${\cal C}=10000$.
\end{enumerate}
Step 0 was carried out once, then steps 1 and 2 repeated 2000 times each for $k=1, m=124$.

\begin{figure}[!htb]
	\centering
	\includegraphics[width=.495\linewidth]
	{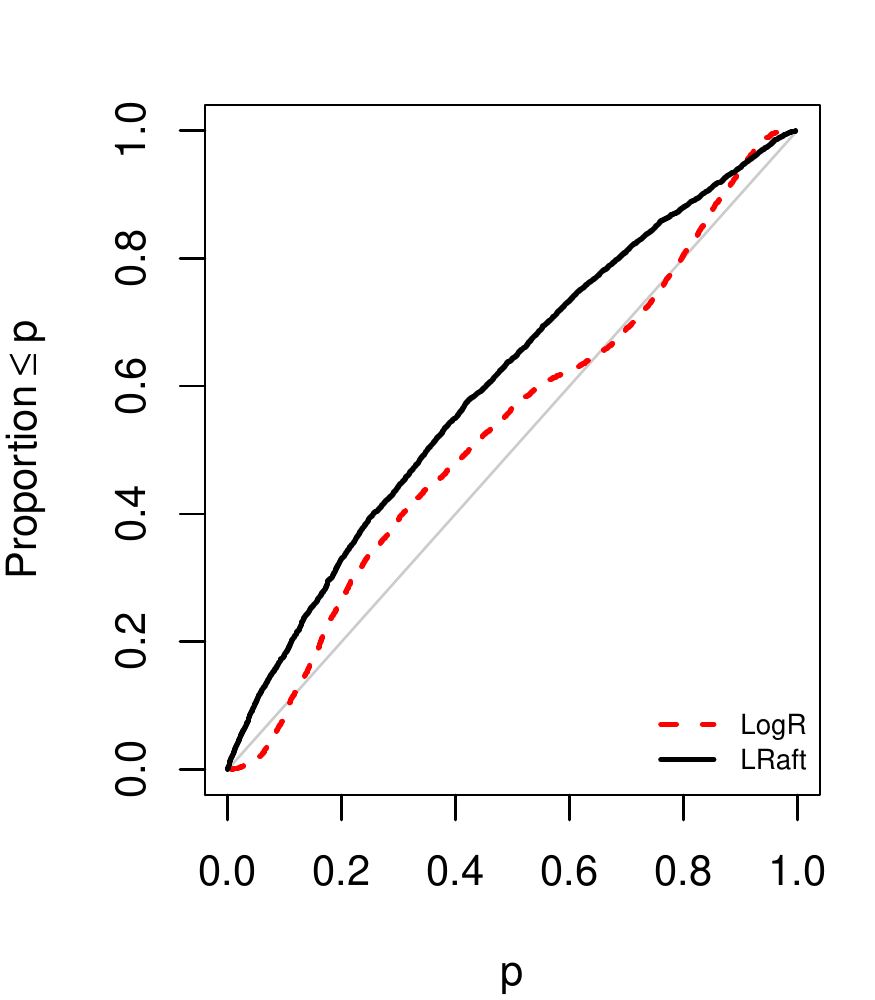}
	\includegraphics[width=.495\linewidth]
	{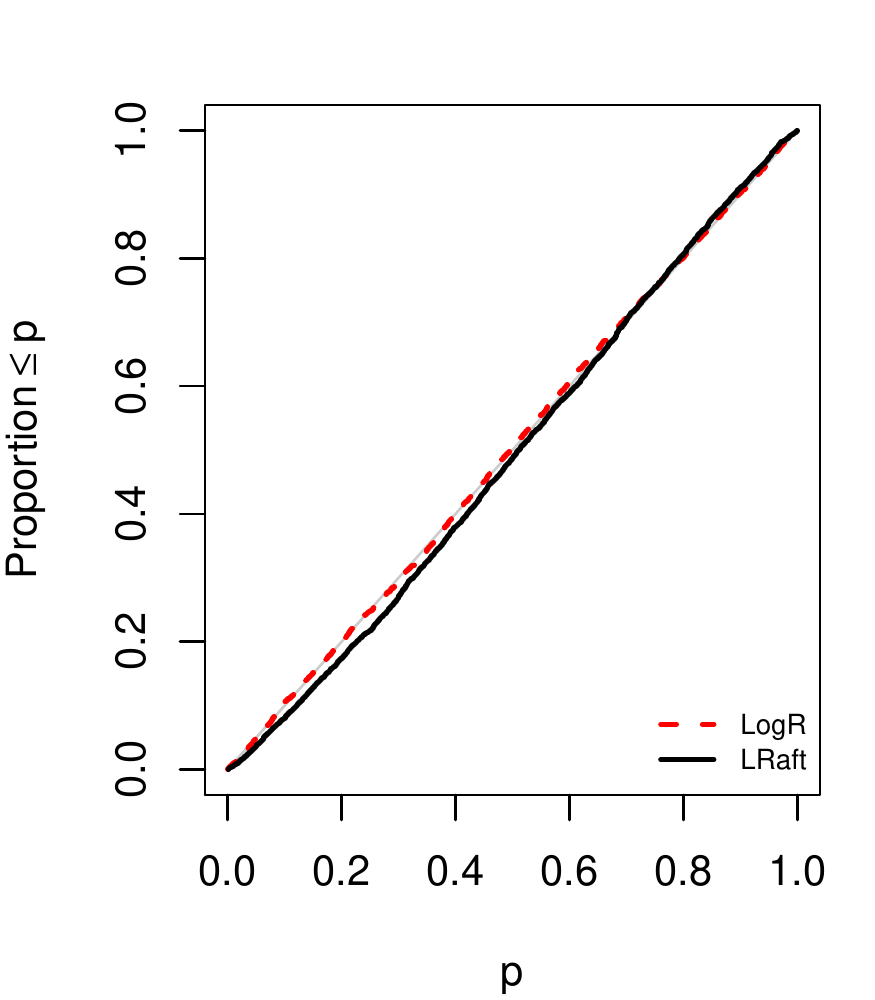}	
	\caption{Empirical cumulative distributions of p-values for the different test procedures described in Section \ref{sect:IPZ}. For the procedure corresponding to the left panel, the failure indicators $D_i$ are held fixed over re-assignments. In contrast, the proposed method corresponding to the right panel allows for the set of censored individuals to vary over re-assignments.
	\label{plot-pvaluesECDF-complete_rand-sims-cens_TRUE-add_FALSE-model_2-n_256-modelH0_2-H0}
	}
	\end{figure}

The empirical cumulative distribution functions (ECDFs) of the LogR and LRaft p-values holding $\bf D$ fixed over re-assignments are plotted in the left panel of
Figure~\ref{plot-pvaluesECDF-complete_rand-sims-cens_TRUE-add_FALSE-model_2-n_256-modelH0_2-H0}.
Neither test controlled the nominal type I error rate in general, with both ECDFs above the diagonal indicating inflated rejection rates of $H_0$ above the nominal size. While the empirical type I error rate of the LogR test was below the nominal rate at certain significance levels, this is not guaranteed to be the case in general.
	
The empirical results using the proposed method in Section~\ref{sect:IPZ} are shown in the right panel of 
Figure~\ref{plot-pvaluesECDF-complete_rand-sims-cens_TRUE-add_FALSE-model_2-n_256-modelH0_2-H0}.
The LogR and LRaft tests both had type I error rates 
that approximately equal the nominal size for all significance levels $\alpha$, with both ECDFs lying approximately on the diagonal.
Similar results for other values of $k$ and $m$ are shown in 
Web Figures~\ref{sims-withcens-add-LH-0} and \ref{sims-withcens-add-LH-1}.
The proposed method was further evaluated using a different (symmetric) interference structure that was generated as a linear preferential attachment network following \citet{Jagadeesan2017}. 
Settings where the uniformity failure times were correlated between individuals and were correlated with censoring times were also considered. Details of these studies (32 different simulation settings) are given in Web Appendix~\ref{sect:sims-withcens-add}. The results, displayed in Web Figures~\ref{sims-withcens-add-LH-0} to \ref{sims-withcens-corr-add-PA-1}, demonstrate that the proposed method controlled the type I error at approximately the nominal level over a variety of scenarios.

\begin{figure}[!htb]
\centering
    \includegraphics[width=.49\linewidth]
    {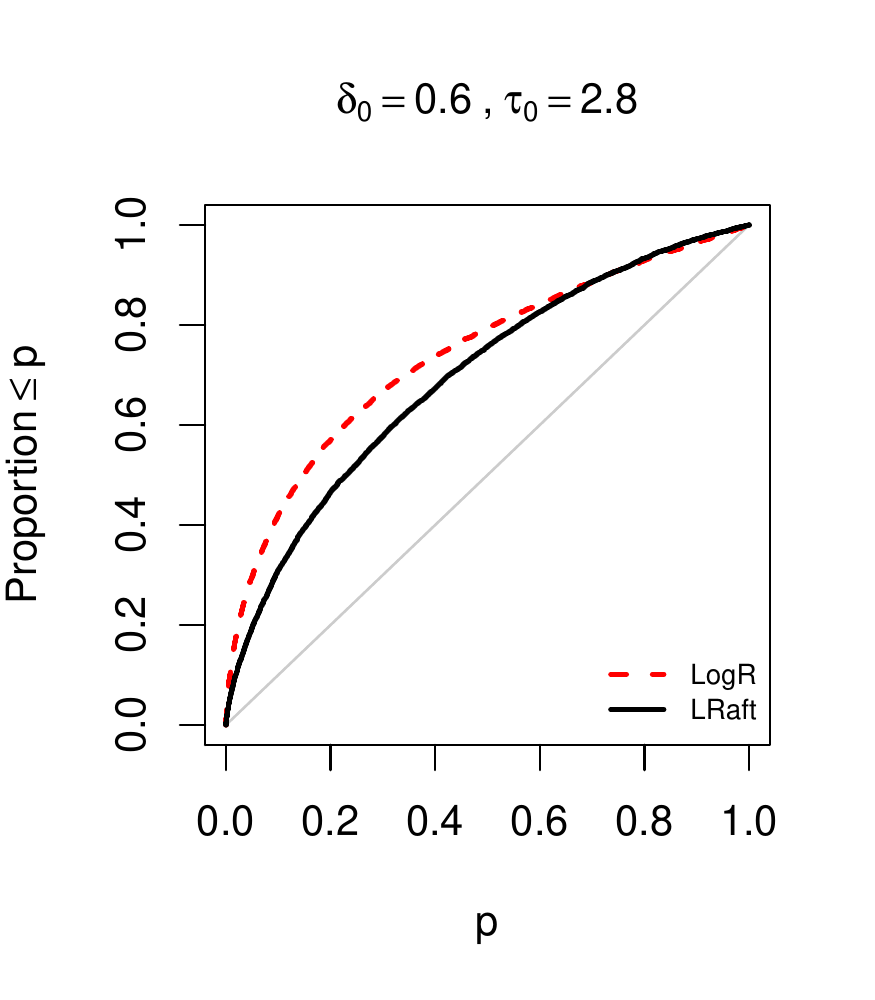}
    \includegraphics[width=.49\linewidth]
	{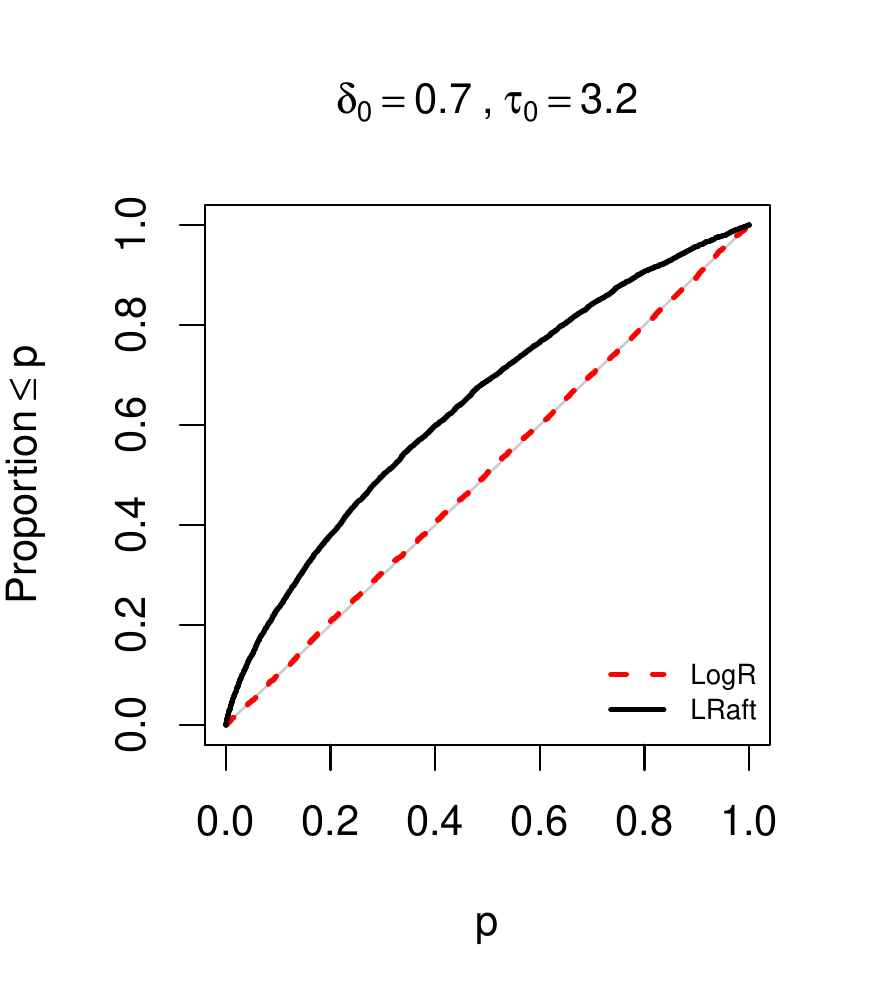}
	\caption{
	Empirical cumulative distributions of p-values from a simulation study described in Web Appendix~\ref{sect:example_n256censoring_power-plots} using the proposed method in Section~\ref{sect:IPZ}. The true parameter values used to generate the data were $(\delta,\tau)=(0.7,2.8)$. The left panel corresponds to testing the null $H_0: (\delta_0,\tau_0)=(0.6,2.8)$, and the right panel corresponds to testing the null $H_0: (\delta_0,\tau_0)=(0.7,3.2)$.
	\label{plot-pvaluesECDF-complete_rand-sims-cens-n_128-m_64-d_07-t_28-}}
\end{figure}

Additional simulation studies were conducted to compare the power of the LRaft and LogR tests. Details of these studies are described in Web Appendix~\ref{sect:example_n256censoring_power-plots}. The results displayed in Figure~\ref{plot-pvaluesECDF-complete_rand-sims-cens-n_128-m_64-d_07-t_28-} correspond to testing the null hypotheses $H_0: (\delta_0, \tau_0)=(0.6,2.8)$ (left panel) and $H_0: (\delta_0, \tau_0)=(0.7,3.2)$ (right panel) when the true data generating parameter values were $(\delta,\tau)=(0.7,2.8)$. Power using the LRaft and LogR tests was similar for $(\delta_0, \tau_0)=(0.6,2.8)$, whereas for $(\delta_0,\tau_0)=(0.7,3.2)$ the LRaft test was more powerful with LogR having power approximately equal to the nominal significance level. The observed lack of power of LogR to detect spillover effects different from that posited under the null aligns with intuition since this statistic only compares (censored) uniformity trial outcomes between treated and untreated individuals, with no attempt to account for the proportion (or number) of treated neighbors. Results for other assumed values of $(\delta_{0}, \tau_{0})$, as well as empirical coverage of the LRaft and LogR 95\% confidence sets, are provided in Web Appendix~\ref{sect:example_n256censoring_power-plots}.

In summary, results from these simulation studies indicate the randomization test procedure in Section~\ref{sect:IPZ} controls the type I error (empirically) over a range of settings, and the LRaft test tends to be as or more powerful than the LogR test. Moreover, the LogR test can lack power to detect spillover effects and thus is not recommended in practice when assuming the additive causal model.

\vspace{-12pt}

\section{Application to randomized trial of cholera vaccine \label{sect:cholera}}

In this section the methods described above are utilized 
to assess the effects of cholera vaccination in a placebo-controlled individually-randomized trial in Matlab, Bangladesh \citep{Ali:2005}.
In prior analyses of these data, \citet{Ali:2005} found a negative association between an individual's risk of cholera infection and the proportion of individuals vaccinated in the area surrounding an individual's residence, suggesting possible interference. 
Similarly, analysis by \citet{emch2009spatial} found that the risk of cholera was inversely related with vaccine coverage in environmental networks that were connected via shared ponds.
Likewise, \citet{Root:2011aa} concluded that the risk of cholera among placebo recipients was inversely associated with level of vaccine coverage in their social networks. 
Motivated by these association analyses, \citet{Perez-Heydrich:2014aa} used inverse probability weighted estimators 
to provide evidence of a significant indirect (spillover) effect of cholera vaccination. 
However, \citeauthor{Perez-Heydrich:2014aa} assumed partial interference based on a spatial clustering of individuals into groups and did not account for right censoring. Misspecification of the interference structure and failure to account for right censoring may bias results. 
The analysis below considers other possible interference structures and allows for right censoring. 

All children aged 2-15 years and females over 15 years 
in the Matlab research site of the International Centre for Diarrheal Disease Research, Bangladesh
were individually assigned randomly to one of three possible treatments:
B subunit-killed whole cell oral cholera vaccine; killed whole cell-only oral cholera vaccine; or Escherichia coli K12 placebo. 
Recipients of either vaccine were grouped together for analysis as the vaccines were 
identical in cellular composition and similar in protective efficacy in previous analyses.
Denote $Z=0$ for those assigned to placebo, and $Z=1$ for those assigned to either vaccine.
Individuals were only included in the analysis if they had completely ingested an initial dose and had completely or almost completely ingested at least one additional dose.
There were a total of $n=72965$ individuals in the randomized trial subpopulation for analysis,
with $m=48660$ assigned to vaccine and $n-m=24305$ to placebo.
The primary outcome for analysis was the (failure) time in days from the 14th day after the vaccination regimen was completed (end of the immunogenic window; \citealt{clemens1988field}),
until a patient was diagnosed with cholera following presentation for treatment of diarrhea.
Failure times for many trial participants were right censored 
either due to outmigration from the field trial area or death prior to the end of the study,
or administrative censoring at the end of the study on June 1, 1986.

\subsection{Interference specifications}

The vaccine trial is analyzed using one of three different specifications of interference in turn. 
Person-to-person transmission of cholera often takes place within the same bari, i.e., geographically clustered households of patrilineally-related individuals. Therefore, for all three specifications, an individual's interference set includes all other individuals residing in the same bari. In other words, all individuals $i,j$ residing in the same bari have $A_{ij}=1$. There are 6423 geographically discrete baris with each individual residing in exactly one bari. Three different specifications are posited regarding how an individual's interference set may also include individuals in different baris. 

The first specification follows the same approach in \citet{Perez-Heydrich:2014aa}.
Baris are partitioned into `neighborhoods' according to a single linkage agglomerative clustering method. %
No interference is assumed between individuals in different neighborhoods and no additional assumptions are imposed regarding the interference structure. 
That is, partial interference is assumed under this specification.
The average number of individuals in each interference set is 419
with an interquartile range (IQR) of 120--631.

\citet{Ali:2005} found an association between the cholera risk for a placebo recipient and the vaccine coverage among individuals living within a 500 meter (m) radius of the placebo recipient. 
Following \citeauthor{Ali:2005}, the second specification of the individual interference sets assumes an individual's potential outcomes may possibly depend on
those living in a different bari within a 500m radius of the bari s/he resided in.
This specification does not assume partial interference.
The average number of individuals in each interference set under this specification is 499 (IQR 339--626). Baris in the same neighborhood under the first specification may be more than 500m apart, e.g., in sparsely populated regions; conversely, baris in different (possibly adjacent) neighborhoods may be less than 500m apart. Hence, $A_{ij}\!=\!1$ under either specification does not imply that $A_{ij}\!=\!1$ under the other specification.

\begin{figure}[!htb]
    \centering
    \begin{minipage}{0.33\textwidth}
        \centering
        \includegraphics[width=\linewidth]{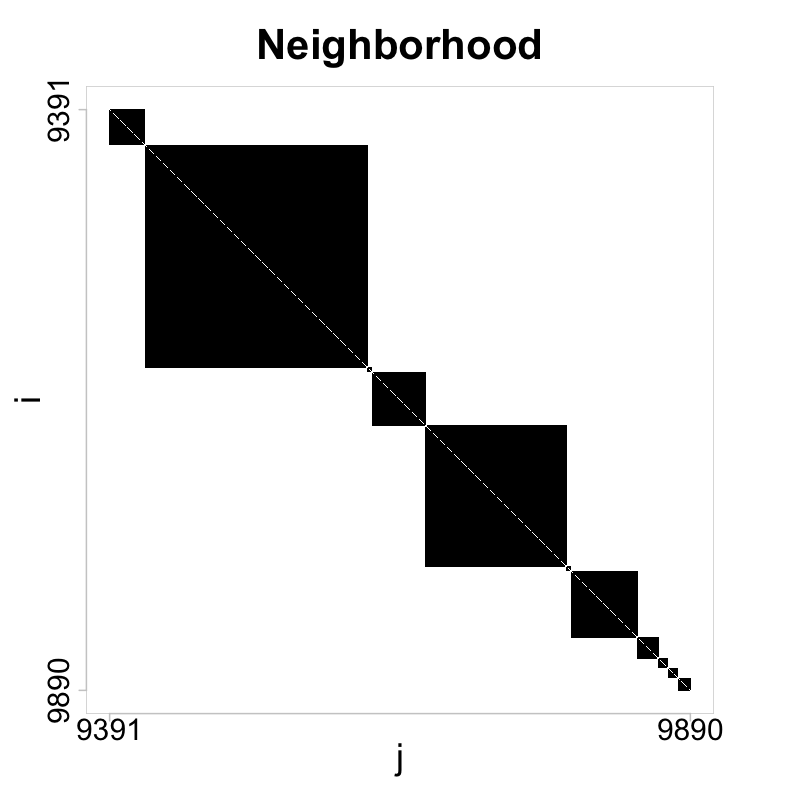}
    \end{minipage}%
    \begin{minipage}{0.33\textwidth}
        \centering
        \includegraphics[width=\linewidth]{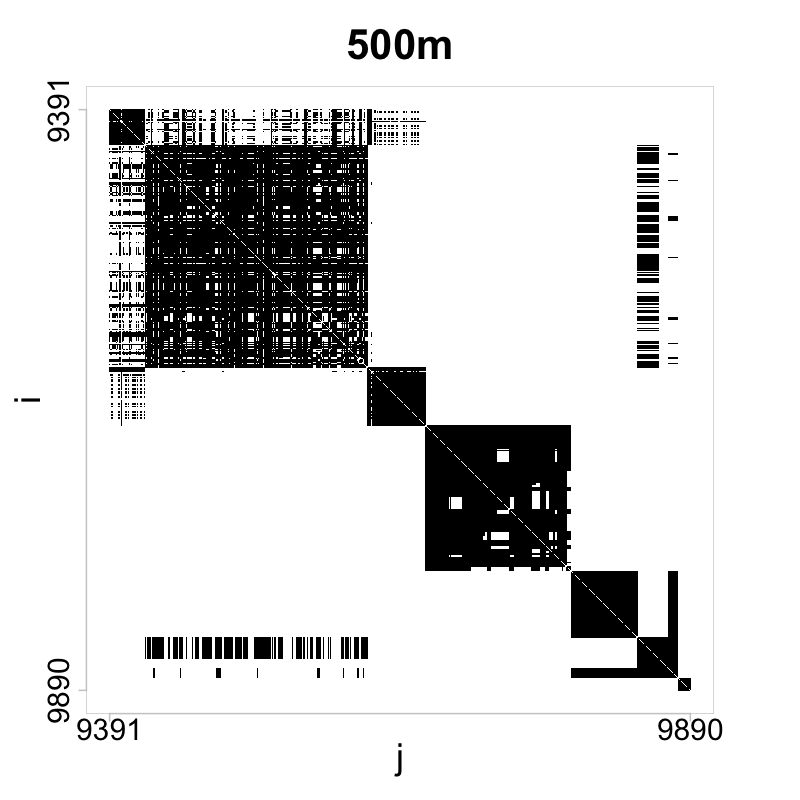}
    \end{minipage}
    \begin{minipage}{0.33\textwidth}
        \centering
        \includegraphics[width=\linewidth]{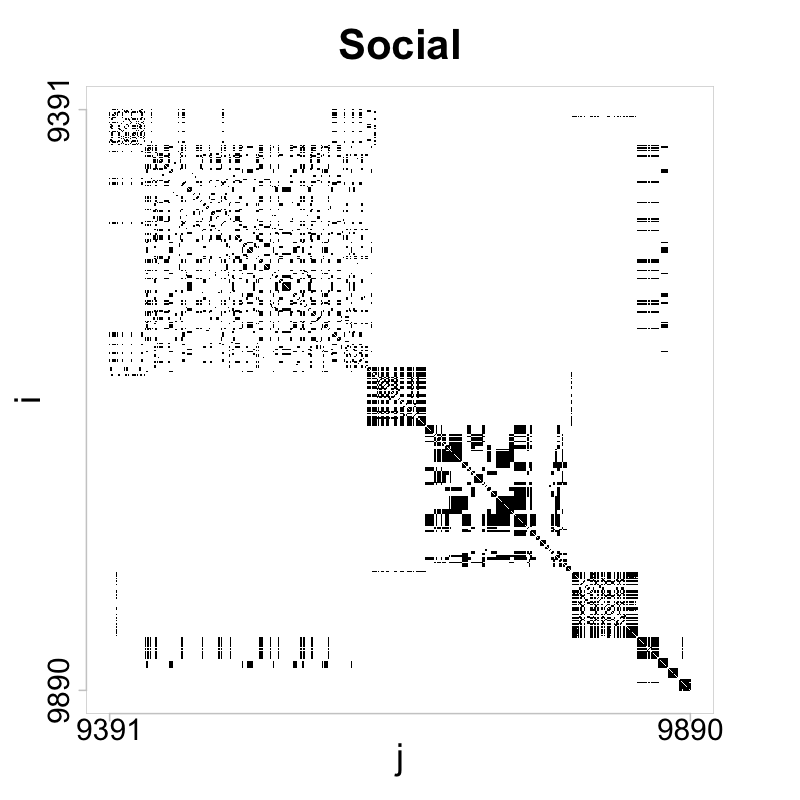}
    \end{minipage}    
	\caption{Submatrices of the interference matrices for 500 selected participants in the randomized cholera vaccine trial, based on Neighborhood (left), 500m (center), and Social (right) interference specifications. \label{fig:cholera_adjmtx_defs}}
	\end{figure}

\begin{figure}[!htb]
    \centering
    \begin{minipage}{0.33\textwidth}
        \centering
        \includegraphics[width=\linewidth]{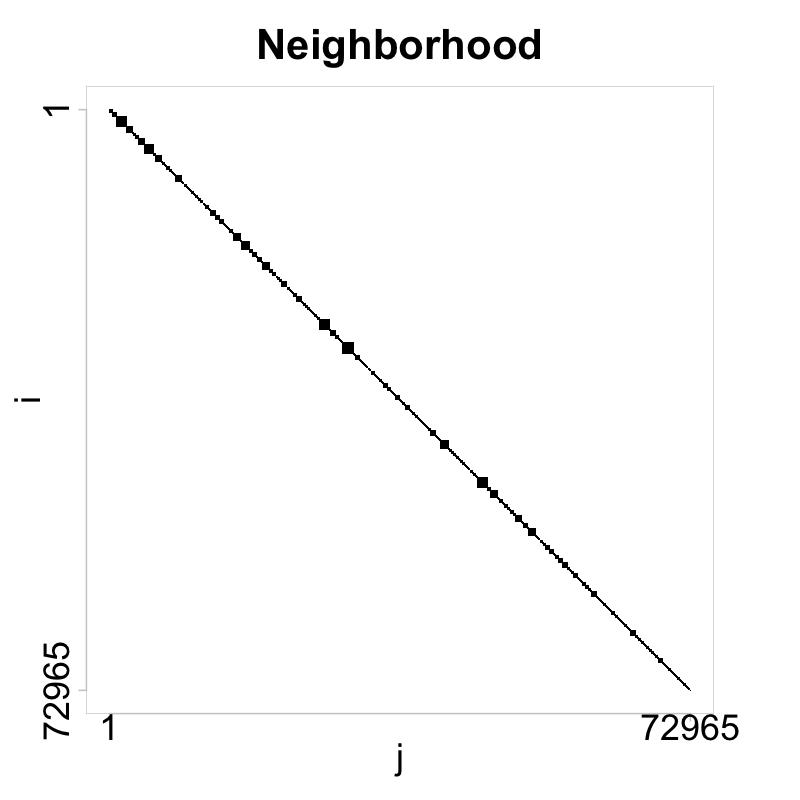}
    \end{minipage}%
    \begin{minipage}{0.33\textwidth}
        \centering
        \includegraphics[width=\linewidth]{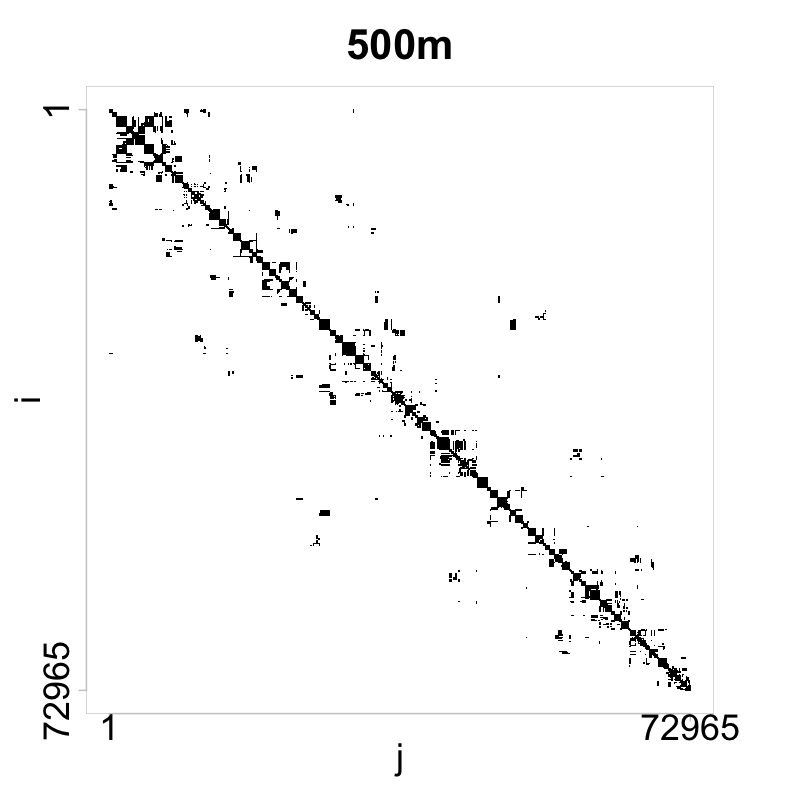}
    \end{minipage}
    \begin{minipage}{0.33\textwidth}
        \centering
        \includegraphics[width=\linewidth]{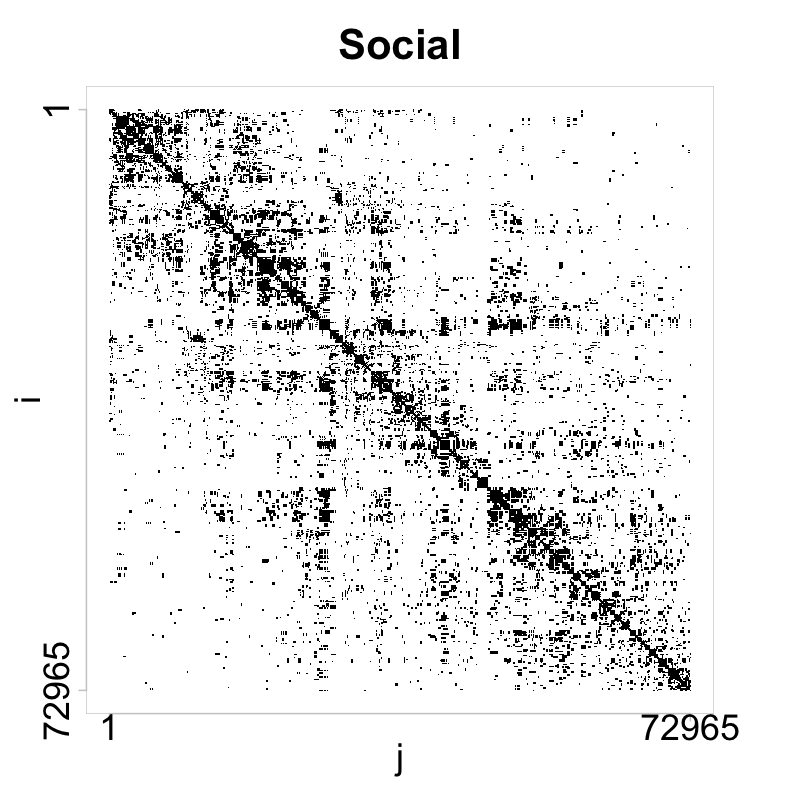}
    \end{minipage}    
	\caption{Interference matrices for all $n=72965$ participants in the randomized cholera vaccine trial, based on Neighborhood (left), 500m (center), and Social (right) interference specifications. \label{fig:cholera_adjmtx_defs-all}}
	\end{figure}

The previous two specifications assume a local interference structure based on geographical location of individuals' households.
Following \citet{Root:2011aa}, the third interference structure is 
defined according to a kinship-based social network between baris.
The Matlab Demographic Surveillance System recorded the exact dates and bari of residence over time for each individual. An individual who migrated between two baris, primarily due to kinship relationships such as marriage, created a non-directional social tie between the baris.
The average number of individuals in each interference set under this specification is 162 (IQR 70--225).
Submatrices of the interference matrices for 500 selected participants under each of the three specifications (`Neighborhood,' `500m,' and `Social') are depicted in Figure~\ref{fig:cholera_adjmtx_defs}.
The interference matrices for all $n=72965$ participants are shown in Web Figure~\ref{fig:cholera_adjmtx_defs-all}.

	The study population also included 44887 individuals who did not participate in the randomized trial,
	and thus have zero probability of receiving either cholera vaccine.
	However, most of these individuals also resided in the same baris as those
	who took part in the trial: 
	5661 baris contained a mixture of participants and non-participants,
	with a median participation rate of 71\% within a bari.
	Since the three specified interference sets are defined based on baris,
	the definition of $G_i$ is expanded
	to include non-trial participants as follows.
	Let $B_i$ be the total number in the study population,
	regardless of trial participation, who may possibly interfere with person $i$,
	so that $B_i \geq A_i \geq T_i$.
	Denote the proportion of $B_i$ who receive treatment as
	$G_i^{\ast}$, i.e., $G_i^{\ast} = {T_i}/{B_i}$.

\subsection{Results}

\begin{figure}[!htb]
    \centering
    \begin{minipage}{{0.33}\textwidth}
        \centering
        \includegraphics[width=\linewidth]{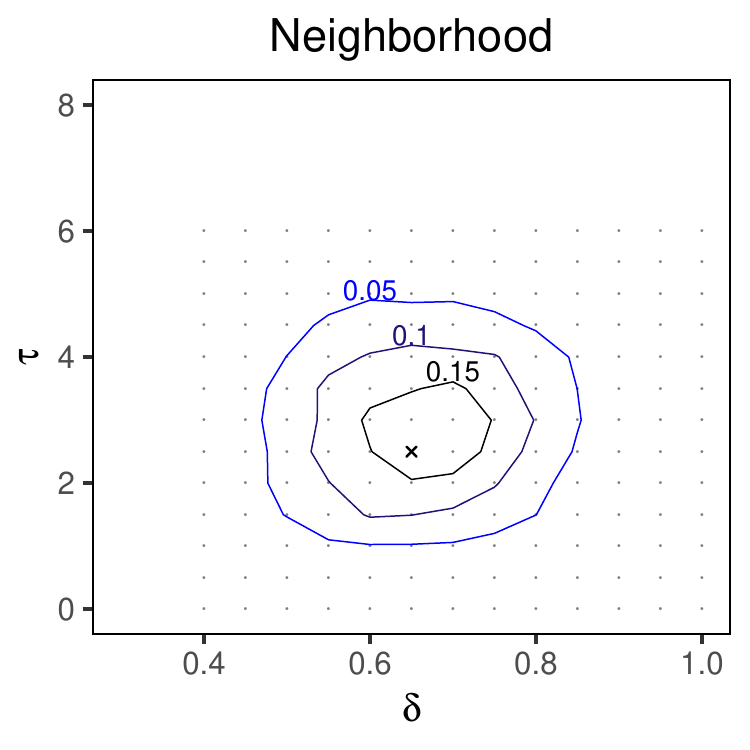}
    \end{minipage}%
    \begin{minipage}{{0.33}\textwidth}
        \centering
        \includegraphics[width=\linewidth]{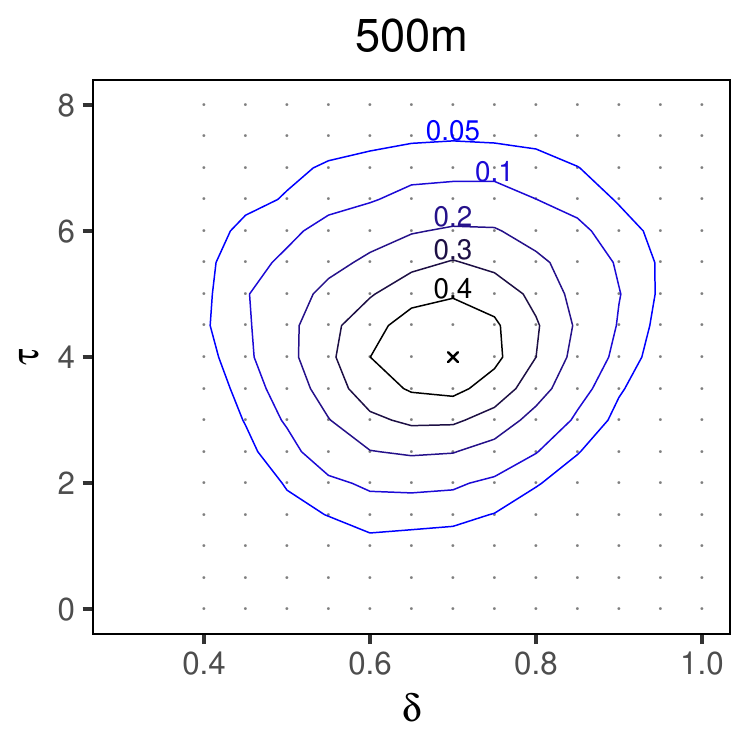}
    \end{minipage}
    \begin{minipage}{{0.33}\textwidth}
        \centering
        \includegraphics[width=\linewidth]{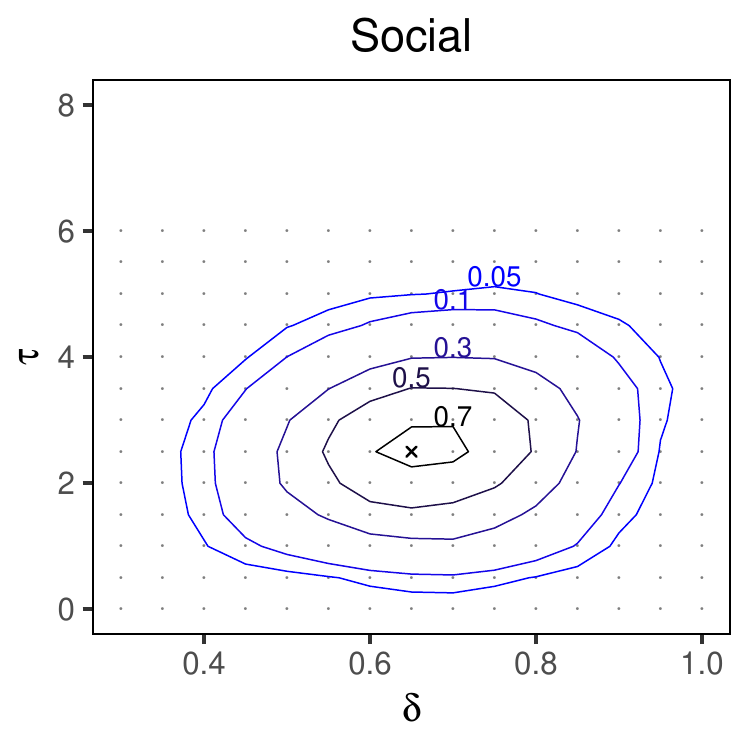}
    \end{minipage}
	\caption{LRaft 95\% confidence sets for ($\delta,\tau$)
	under the additive model ${\mathcal F}_{\rm add}$, and each
	specified interference matrix (Neighborhood, 500m or Social) for the cholera data.
	The contours indicate values of ($\delta_0,\tau_0$) yielding the same p-values, with darker hues indicating larger p-values.
	The boundaries of the 95\% confidence set are demarcated by the contour lines that indicate p-values of at least 0.05. 
	The point estimate ($\hat \delta, \hat \tau$) corresponding to the highest p-value 
	under each interference structure is indicated by \texttt{x}.
	\label{fig:cholera_ci_model2}}
	\end{figure}

For each specified interference matrix,
confidence sets for ($\delta,\tau$) were constructed 
under the causal model $Y_i = y_i({\bf 0}) \exp (\delta Z_i + \tau G_i^{\ast})$
by conducting hypothesis tests over a discrete grid of values of $(\delta_0, \tau_0)$.
It was not computationally feasible to enumerate $\Omega$ exactly
with ${72965 \choose 48660} \approx 10^{20162}$ possible re-assignments,
so p-values were calculated
with ${\cal C}=4000$ random draws from $\Omega$.
The LRaft 95\% confidence sets are plotted in Figure~\ref{fig:cholera_ci_model2},
with the contours indicating ($\delta_0,\tau_0$) values yielding the same p-values.
The boundaries of the 95\% confidence set are demarcated
by the contour lines that indicate p-values at least as large as 0.05.

There is evidence vaccination has an effect on the risk of cholera 
as the 95\% confidence sets exclude $(\delta,\tau)=(0,0)$ 
under all three interference specifications. 
Point estimates of the joint treatment effects,
corresponding to values of ($\delta_0,\tau_0$) with the largest p-value, are positive,
suggesting the effect of the vaccine in reducing the risk of cholera
is a combination of protective direct and spillover effects.
The direct effect estimates are similar across the three interference specifications, whereas the spillover effect estimate is somewhat higher for the Social interference specification. 
For the 500m interference structure,
the estimated treatment effect is
$(\hat \delta, \hat \tau)=(0.7, 4.0)$.
We offer two interpretations of $(\hat \delta, \hat \tau)$
under the additive causal model.
First, the average time until cholera diagnosis had everyone not received vaccine 
(i.e., the uniformity trial) is estimated to be $\exp(0.7+4.0) = \exp(4.7) \approx 110$ times faster than 
if everyone had received vaccine (e.g., the `blanket coverage' trial).
Second, 
the estimated risk of cholera incidence at 365 days
under the uniformity trial would be approximately 2.30\%
compared to 0.06\% under the blanket coverage trial, 
corresponding to a 98\% reduction.
The individual parameter estimates also have a straightforward interpretation. For example, holding the proportion of neighbors treated fixed, $\exp(\hat \delta) = \exp(0.7) \approx 2$ is the estimated ratio of survival times when an individual receives treatment versus control. Similarly, holding individual treatment fixed, $\exp(\hat \tau) = \exp(4.0) \approx 55$ is the estimated ratio of survival times when all neighbors are treated compared to no neighbors being treated. 

The BFP model was considered unrealistic a priori for this example because there was no plausible scientific rationale for limiting the spillover effect to those who do not receive the vaccine, and to be strictly smaller in magnitude than the direct effect. Nonetheless, for completeness, inference was carried out for parameters under an assumed BFP model. No p-values were above 0.05, suggesting that the BFP model is a poor fit to the data.

\vspace{-12pt}

\section{Discussion}

In this paper we proposed randomization-based methods 
for assessing the effect of treatment on right-censored outcomes 
in the presence of general interference. 
There are several avenues of possible future related research. 
The adapted $\rm IP_{Z}$ procedure as implemented only allows for unequal censoring based on $Z$. A  proportional hazards model 
may be used in place of the group-specific KM estimators to allow for censoring to differ based on $Z$ and $G$.
Building on the empirical results in this paper, future research could examine theoretical properties of the proposed procedures, e.g., determine conditions under which type I error rate control is guaranteed.
Joint parametric causal models for both the failure times and censoring times in the presence of general interference might also be considered. 
Since inference is contingent on the choice of interference structure assumed,
possible extensions include developing sensitivity analysis methods for assessing robustness to interference structure misspecification. Alternatively, extensions of randomization-based inference approaches that do not require a parametric causal model, such as \citet{savje2017average}, to the setting where outcomes are censored could be considered. Methods such as \citet{Athey:2016aa} and \citet{Jagadeesan2017} that use restricted randomizations to improve statistical power and computational speed might also be considered.
While illustrated in this paper using data from an individually-randomized trial, the proposed methods can be employed in cluster-randomized trials. 
Finally, although this paper has focused on two specific causal models,
the proposed methods are general and easily extended to other causal models.
			
\backmatter

\section*{Acknowledgements}
The authors thank Brian Barkley, Sujatro Chakladar, Bradley Saul, the Editor, Associate Editor and two reviewers for helpful comments.
This research was supported by NIH grant R01 AI085073-05 and by a Gillings Innovation Laboratory award from the UNC Gillings School of Global Public Health. 
Computational resources and services were provided by the VSC (Flemish Supercomputer Center), funded by the Research Foundation Flanders (FWO) and the Flemish Government -- department EWI.
The content is solely the responsibility of the authors and does not 
represent the official views of the National Institutes of Health.

\vspace*{-8pt}

\bibliographystyle{biom}
\bibliography{bibinterfere}

\begin{thebibliography}{}

\bibitem[\protect\citeauthoryear{Ali, Emch, von Seidlein, Yunus, Sack, Rao,
  Holmgren, and Clemens}{Ali et~al.}{2005}]{Ali:2005}
Ali, M., Emch, M., von Seidlein, L., Yunus, M., Sack, D.~A., Rao, M., Holmgren,
  J., and Clemens, J.~D. (2005).
\newblock Herd immunity conferred by killed oral cholera vaccines in
  bangladesh: a reanalysis.
\newblock {\em The Lancet} {\bf 366,} 44--49.

\bibitem[\protect\citeauthoryear{Athey, Eckles, and Imbens}{Athey
  et~al.}{2018}]{Athey:2016aa}
Athey, S., Eckles, D., and Imbens, G.~W. (2018).
\newblock Exact p-values for network interference.
\newblock {\em Journal of the American Statistical Association} {\bf 113,}
  230--240.

\bibitem[\protect\citeauthoryear{Bowers, Fredrickson, and Aronow}{Bowers
  et~al.}{2016}]{Bowers:2016aa}
Bowers, J., Fredrickson, M.~M., and Aronow, P.~M. (2016).
\newblock Research note: A more powerful test statistic for reasoning about
  interference between units.
\newblock {\em Political Analysis} {\bf 24,} 395--403.

\bibitem[\protect\citeauthoryear{Bowers, Fredrickson, and Panagopoulos}{Bowers
  et~al.}{2012}]{Bowers:2012aa}
Bowers, J., Fredrickson, M.~M., and Panagopoulos, C. (2012).
\newblock Reasoning about interference between units: A general framework.
\newblock {\em Political Analysis} {\bf 21,} 97--124.

\bibitem[\protect\citeauthoryear{Clemens, Harris, Sack, Chakraborty, Ahmed,
  Stanton, Khan, Kay, Huda, Khan, et~al\mbox{.}}{Clemens
  et~al.}{1988}]{clemens1988field}
Clemens, J.~D., Harris, J.~R., Sack, D.~A., Chakraborty, J., Ahmed, F.,
  Stanton, B.~F., Khan, M.~U., Kay, B.~A., Huda, N., Khan, M., et~al. (1988).
\newblock Field trial of oral cholera vaccines in {B}angladesh: results of one
  year of follow-up.
\newblock {\em Journal of Infectious Diseases} {\bf 158,} 60--69.

\bibitem[\protect\citeauthoryear{Collett}{Collett}{2003}]{Collett:2003aa}
Collett, D. (2003).
\newblock {\em Modelling {S}urvival {D}ata in {M}edical {R}esearch}.
\newblock Chapman \& Hall/CRC, {S}econd edition.

\bibitem[\protect\citeauthoryear{Csardi and Nepusz}{Csardi and
  Nepusz}{2006}]{igraph2006}
Csardi, G. and Nepusz, T. (2006).
\newblock The igraph software package for complex network research.
\newblock {\em InterJournal, Complex Systems} {\bf 1695,} 1--9.

\bibitem[\protect\citeauthoryear{Emch, Ali, Root, and Yunus}{Emch
  et~al.}{2009}]{emch2009spatial}
Emch, M., Ali, M., Root, E.~D., and Yunus, M. (2009).
\newblock Spatial and environmental connectivity analysis in a cholera vaccine
  trial.
\newblock {\em Social Science \& Medicine} {\bf 68,} 631--637.

\bibitem[\protect\citeauthoryear{Halloran and Hudgens}{Halloran and
  Hudgens}{2016}]{Halloran:2016aa}
Halloran, M.~E. and Hudgens, M.~G. (2016).
\newblock Dependent happenings: a recent methodological review.
\newblock {\em Current Epidemiology Reports} {\bf 3,} 297--305.

\bibitem[\protect\citeauthoryear{Hudgens and Halloran}{Hudgens and
  Halloran}{2008}]{Hudgens:2008aa}
Hudgens, M.~G. and Halloran, M.~E. (2008).
\newblock Toward causal inference with interference.
\newblock {\em Journal of the American Statistical Association} {\bf 103,}
  832--842 

\bibitem[\protect\citeauthoryear{Jagadeesan, Pillai, and Volfovsky}{Jagadeesan
  et~al.}{2017}]{Jagadeesan2017}
Jagadeesan, R., Pillai, N., and Volfovsky, A. (2017).
\newblock {Designs for estimating the treatment effect in networks with
  interference}.
\newblock {\em arXiv preprint arXiv:1705.08524} .

\bibitem[\protect\citeauthoryear{Lehmann and Romano}{Lehmann and
  Romano}{2005}]{lehmann2005testing}
Lehmann, E.~L. and Romano, J.~P. (2005).
\newblock {\em Testing {S}tatistical {H}ypotheses}.
\newblock Springer Texts in Statistics. Springer New York, {T}hird edition.

\bibitem[\protect\citeauthoryear{Ogburn, Sofrygin, Diaz, and van~der
  Laan}{Ogburn et~al.}{2017}]{ogburn2017causal}
Ogburn, E.~L., Sofrygin, O., Diaz, I., and van~der Laan, M.~J. (2017).
\newblock Causal inference for social network data.
\newblock {\em arXiv preprint arXiv:1705.08527} .

\bibitem[\protect\citeauthoryear{Perez-Heydrich, Hudgens, Halloran, Clemens,
  Ali, and Emch}{Perez-Heydrich et~al.}{2014}]{Perez-Heydrich:2014aa}
Perez-Heydrich, C., Hudgens, M.~G., Halloran, M.~E., Clemens, J.~D., Ali, M.,
  and Emch, M.~E. (2014).
\newblock Assessing effects of cholera vaccination in the presence of
  interference.
\newblock {\em Biometrics} {\bf 70,} 731--741.

\bibitem[\protect\citeauthoryear{Root, Giebultowicz, Ali, Yunus, and Emch}{Root
  et~al.}{2011}]{Root:2011aa}
Root, E.~D., Giebultowicz, S., Ali, M., Yunus, M., and Emch, M. (2011).
\newblock The role of vaccine coverage within social networks in cholera
  vaccine efficacy.
\newblock {\em PloS {ONE}} {\bf 6,} e22971 

\bibitem[\protect\citeauthoryear{Rosenbaum}{Rosenbaum}{2002}]{Rosenbaum:2002}
Rosenbaum, P.~R. (2002).
\newblock {\em Observational {S}tudies}.
\newblock New York : Springer, New York.

\bibitem[\protect\citeauthoryear{Rosenbaum}{Rosenbaum}{2007}]{Rosenbaum:2007aa}
Rosenbaum, P.~R. (2007).
\newblock Interference between units in randomized experiments.
\newblock {\em Journal of the American Statistical Association} {\bf 102,}
  191--200.

\bibitem[\protect\citeauthoryear{S{\"a}vje, Aronow, and Hudgens}{S{\"a}vje
  et~al.}{2017}]{savje2017average}
S{\"a}vje, F., Aronow, P.~M., and Hudgens, M.~G. (2017).
\newblock Average treatment effects in the presence of unknown interference.
\newblock {\em arXiv preprint arXiv:1711.06399} .

\bibitem[\protect\citeauthoryear{Sobel}{Sobel}{2006}]{Sobel:2006aa}
Sobel, M.~E. (2006).
\newblock What do randomized studies of housing mobility demonstrate? {C}ausal
  inference in the face of interference.
\newblock {\em Journal of the American Statistical Association} {\bf 101,}
  1398--1407 

\bibitem[\protect\citeauthoryear{Sussman and Airoldi}{Sussman and
  Airoldi}{2017}]{sussman2017elements}
Sussman, D.~L. and Airoldi, E.~M. (2017).
\newblock Elements of estimation theory for causal effects in the presence of
  network interference.
\newblock {\em arXiv preprint arXiv:1702.03578} .

\bibitem[\protect\citeauthoryear{Wang, Lagakos, and Gray}{Wang
  et~al.}{2010}]{Wang:2010aa}
Wang, R., Lagakos, S.~W., and Gray, R.~J. (2010).
\newblock Testing and interval estimation for two-sample survival comparisons
  with small sample sizes and unequal censoring.
\newblock {\em Biostatistics} {\bf 11,} 676--692.

\end{thebibliography}


\clearpage

\begin{center}
	\textsc{Web Appendices}
\end{center}

\setcounter{section}{0}
\renewcommand\thesection{\Alph{section}}
\renewcommand\thesubsection{\thesection\arabic{subsection}}

\section{Additional type I error simulation results \label{sect:sims-withcens-add}}

\begin{figure}[!htb]
	\centering
	\includegraphics[width=\linewidth]{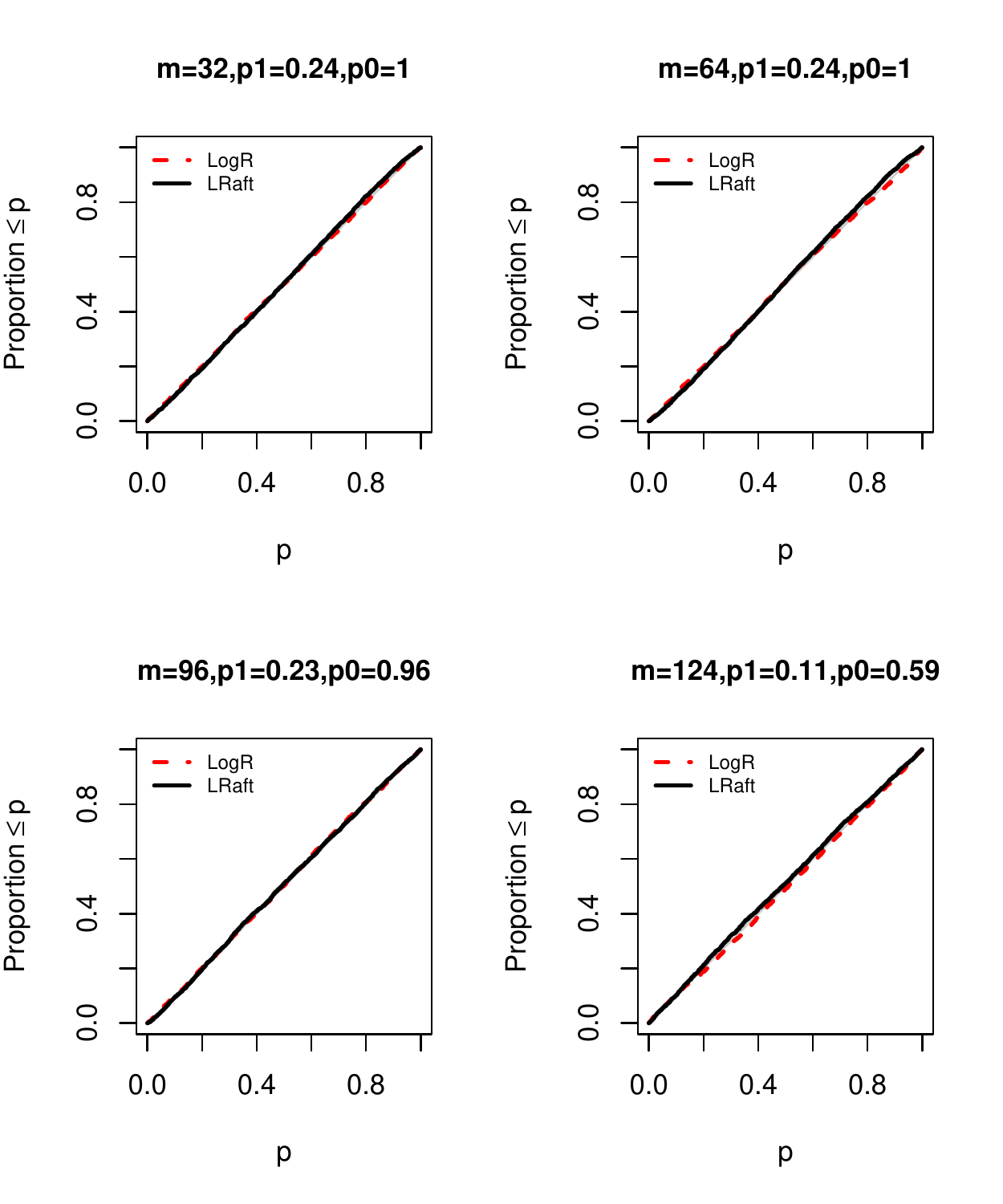}	
	\caption{Empirical cumulative distributions of p-values using the interference structure described in Section~\ref{sect:example_n256censoring_power} and $k=0.6$. Each panel corresponds to a different value of $m$ as stated in the title. The average proportions of observed failures in the $Z=1$ and $Z=0$ groups are stated in the title as `p1' and `p0' respectively.
	\label{sims-withcens-add-LH-0}
	}
	\end{figure}

\begin{figure}[!htb]
	\centering	
	\includegraphics[width=\linewidth]{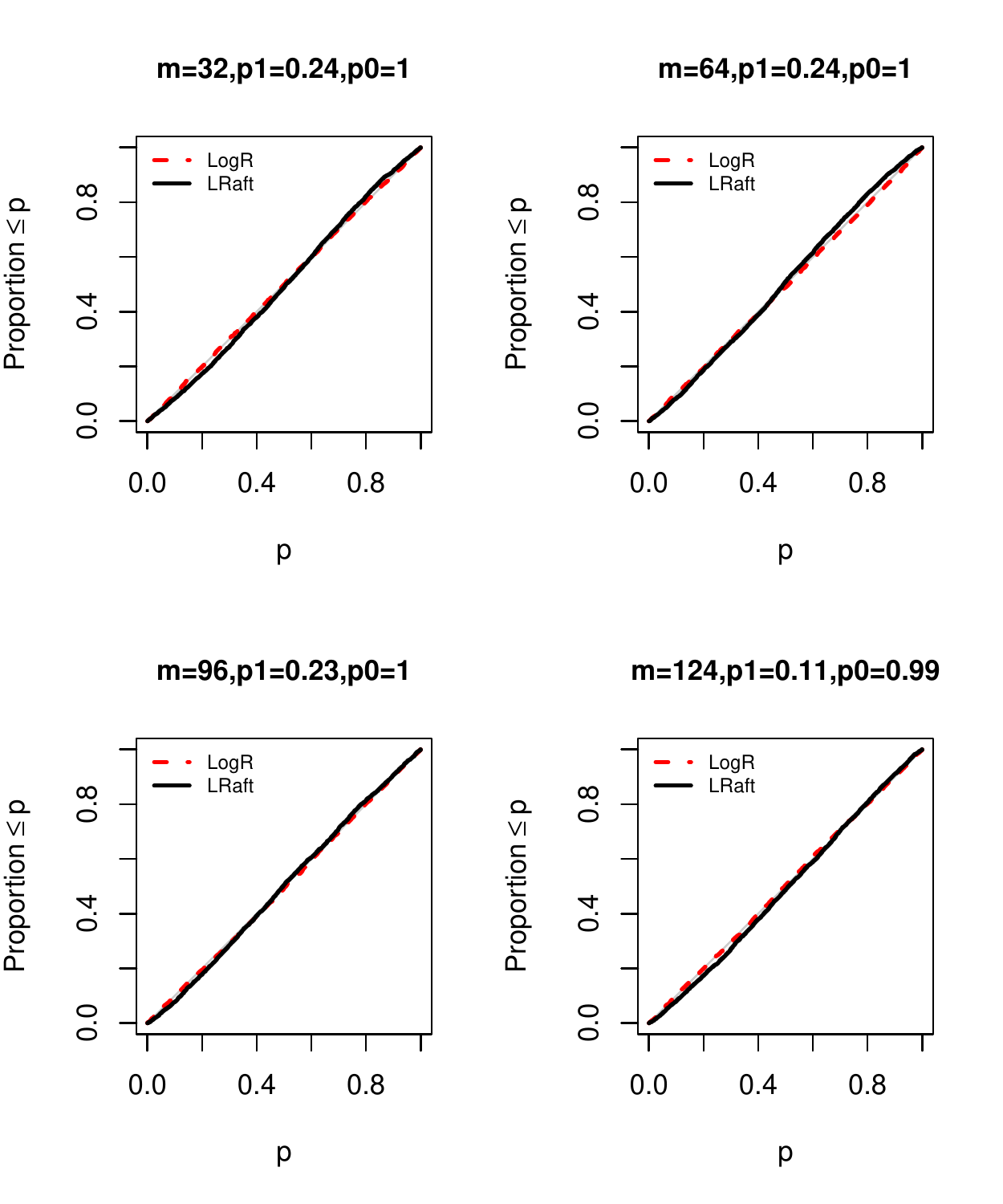}	
	\caption{Empirical cumulative distributions of p-values using the interference structure described in Section~\ref{sect:example_n256censoring_power} and $k=1$. Each panel corresponds to a different value of $m$ as stated in the title. The average proportions of observed failures in the $Z=1$ and $Z=0$ groups are stated in the title as `p1' and `p0' respectively. The bottom right panel is presented in the main paper as the right panel in 
Figure~\ref{plot-pvaluesECDF-complete_rand-sims-cens_TRUE-add_FALSE-model_2-n_256-modelH0_2-H0}.
	\label{sims-withcens-add-LH-1}
	}	
	\end{figure}

The simulation study in Section~\ref{sect:example_n256censoring_power} was repeated for all combinations of $k=0.6,1$ and $m=124,96,64,32$. The empirical cumulative distribution functions (ECDFs) of the LogR and LRaft p-values are plotted in Figures~\ref{sims-withcens-add-LH-0} ($k=0.6$) and \ref{sims-withcens-add-LH-1} ($k=1$). The failure rates were between $11\%$ and $24\%$ in the $Z=1$ arm, and between $59\%$ and $100\%$ in the $Z=0$ arm. 


\begin{figure}[!htb]
	\centering
	\includegraphics[width=\linewidth]{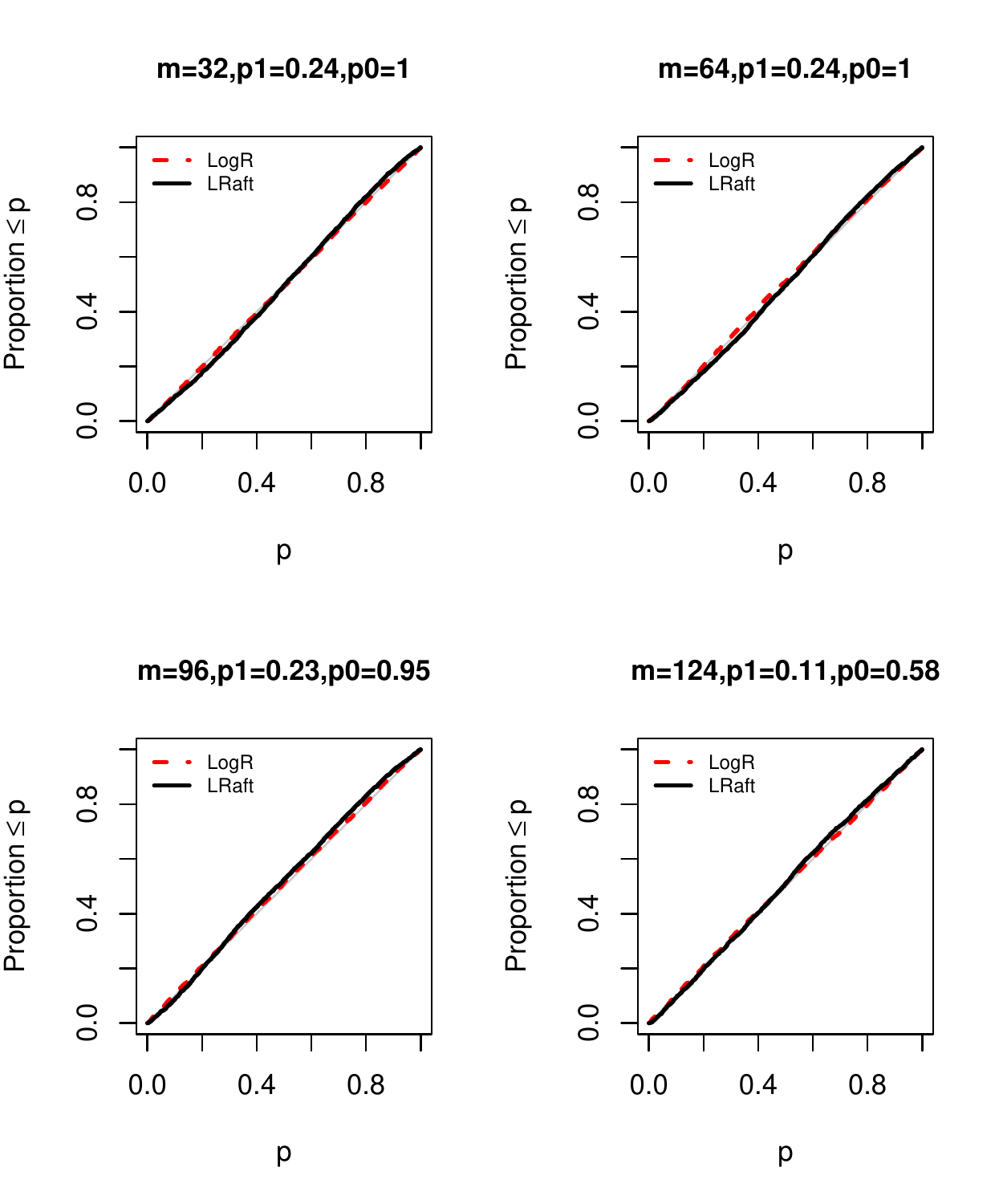}	
	\caption{Empirical cumulative distributions of p-values using the PA interference structure and $k=0.6$. Each panel corresponds to a different value of $m$ as stated in the title. The average proportions of observed failures in the $Z=1$ and $Z=0$ groups are stated in the title as `p1' and `p0' respectively.
	\label{sims-withcens-add-PA-0}
	}
	\end{figure}

\begin{figure}[!htb]
	\centering		
	\includegraphics[width=\linewidth]{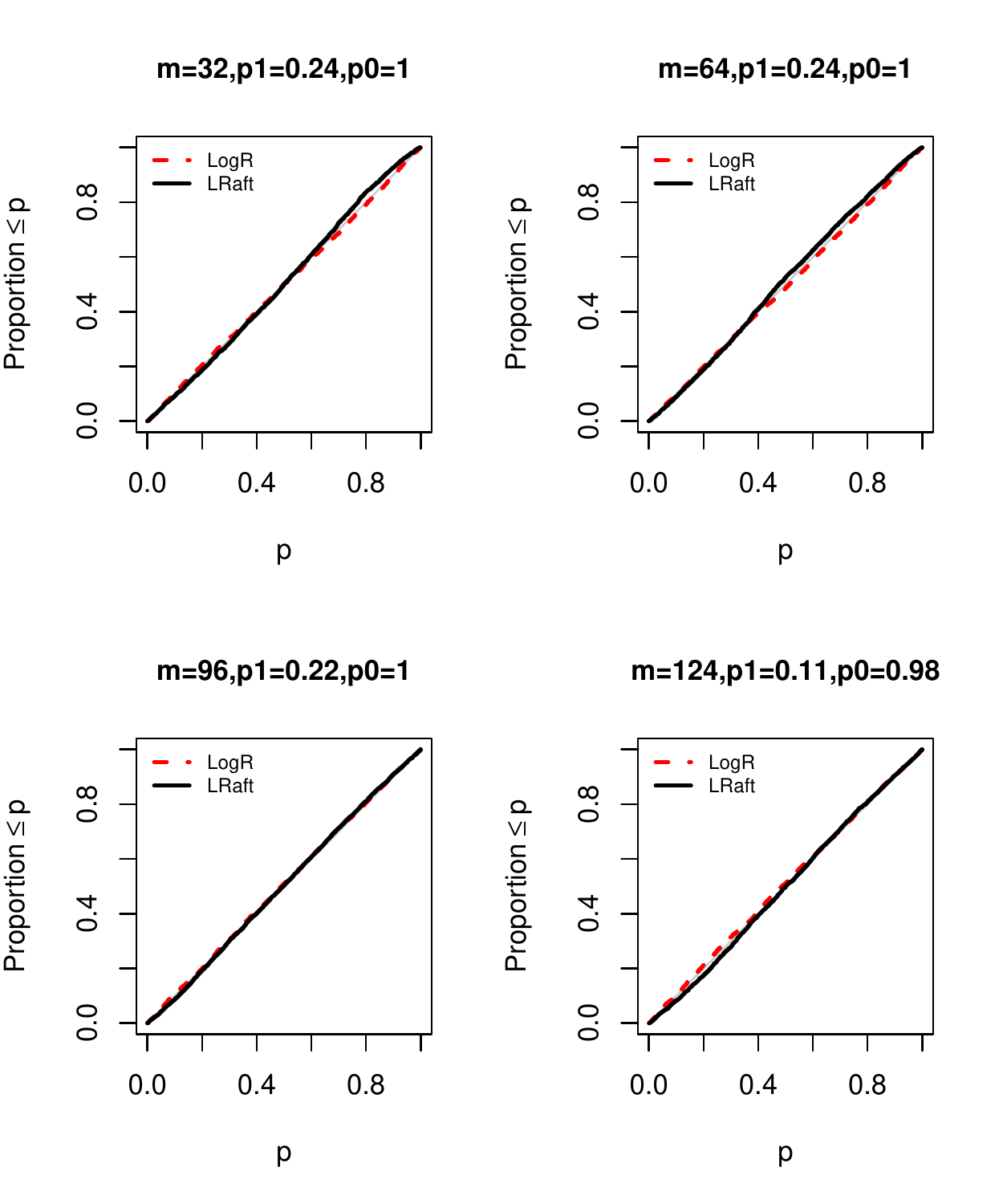}	
	\caption{Empirical cumulative distributions of p-values using the PA interference structure and $k=1$. Each panel corresponds to a different value of $m$ as stated in the title. The average proportions of observed failures in the $Z=1$ and $Z=0$ groups are stated in the title as `p1' and `p0' respectively.
	\label{sims-withcens-add-PA-1}
	}	
	\end{figure}

The simulation study was also carried out using a (symmetric) interference structure that was generated as a linear preferential attachment (PA) network following Section 9.2 of \citet{Jagadeesan2017}. The network is constructed by starting with a single individual and then adding one new individual at a time until there are $n$ individuals in the network. Each new individual that is added, denoted by e.g., $j$, forms an edge with each of the existing individuals $i=1,\ldots,j-1$ with a probability that is proportional to $A_i$; for $j \geq m$, each new individual $j$ that is added forms $m$ new edges. The network can be generated using the \texttt{sample\_pa} function in the \texttt {igraph} package \citep{igraph2006}. 
As in the simulation study in Section~\ref{sect:example_n256censoring_power}, the average number of neighbors is about 16; however, there are now individuals with comparatively large number of neighbors e.g., about 70. This simulation study was then carried out for all combinations of $k=0.6,1$ and $m=124,96,64,32$. Plots of the ECDFs of the p-values are shown in Figures~\ref{sims-withcens-add-PA-0} ($k=0.6$) and \ref{sims-withcens-add-PA-1} ($k=1$). 
The failure rates were between $11\%$ and $24\%$ in the $Z=1$ arm, and between $58\%$ and $100\%$ in the $Z=0$ arm. 


\begin{figure}[!htb]
	\centering
	\includegraphics[width=\linewidth]{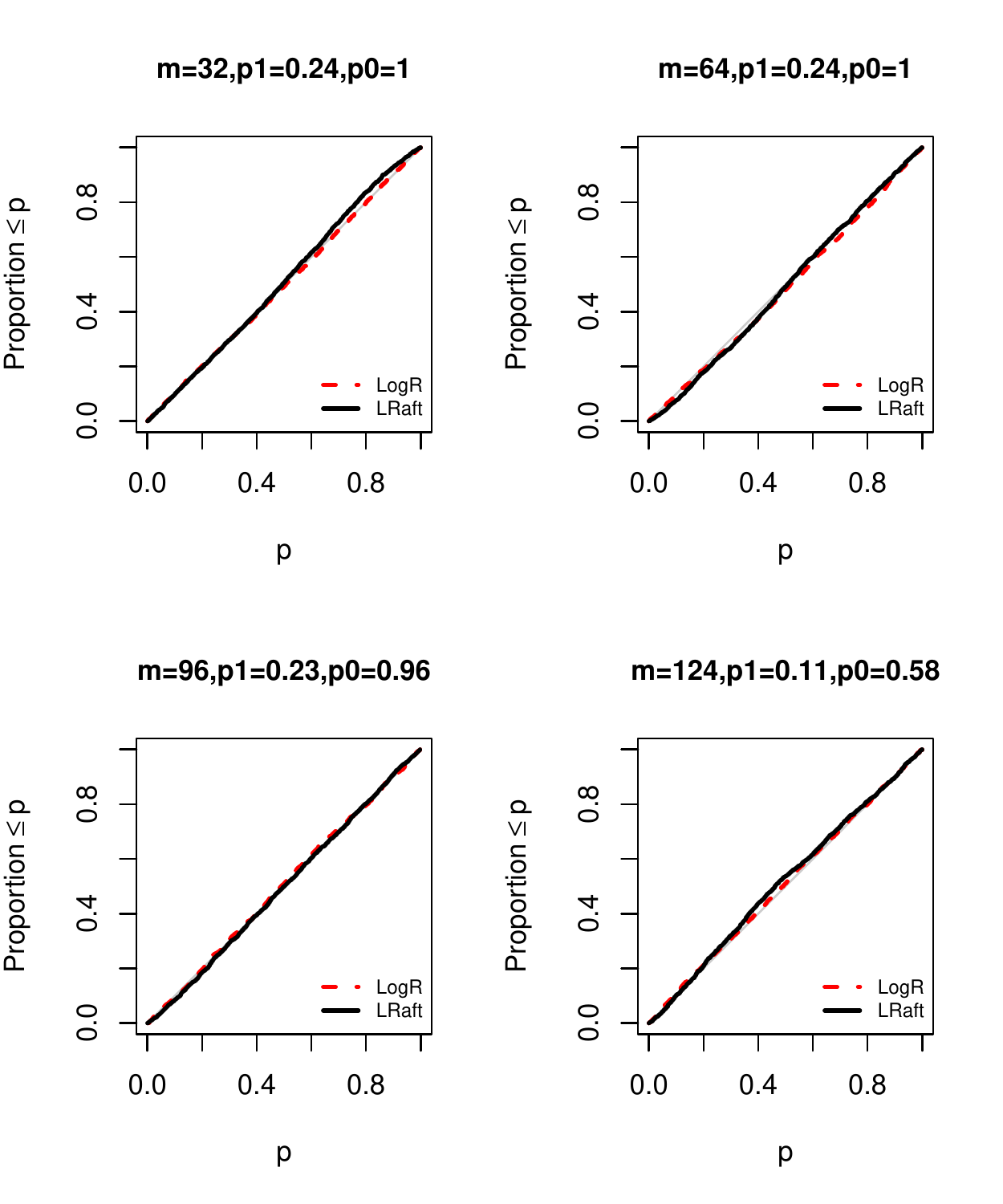}	
	\caption{Empirical cumulative distributions of p-values using the interference structure described in Section~\ref{sect:example_n256censoring_power} and $k=0.6$, with failure times correlated between individuals and with censoring times. Each panel corresponds to a different value of $m$ as stated in the title. The average proportions of observed failures in the $Z=1$ and $Z=0$ groups are stated in the title as `p1' and `p0' respectively.
	\label{sims-withcens-corr-add-LH-0}
	}
	\end{figure}

\begin{figure}[!htb]
	\centering	
	\includegraphics[width=\linewidth]{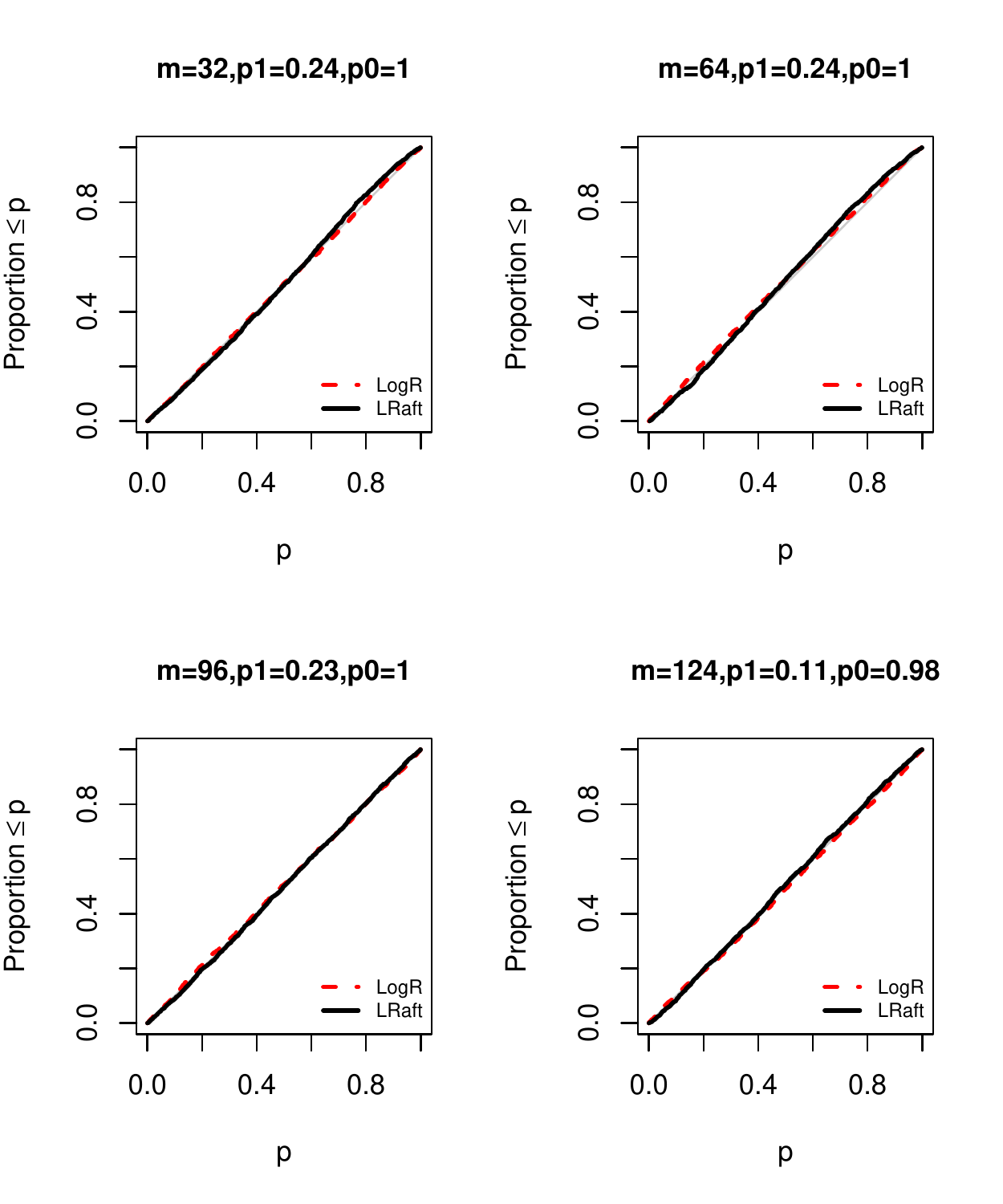}	
	\caption{Empirical cumulative distributions of p-values using the interference structure described in Section~\ref{sect:example_n256censoring_power} and $k=1$, with failure times correlated between individuals and with censoring times. Each panel corresponds to a different value of $m$ as stated in the title. The average proportions of observed failures in the $Z=1$ and $Z=0$ groups are stated in the title as `p1' and `p0' respectively.
	\label{sims-withcens-corr-add-LH-1}
	}	
	\end{figure}

\begin{figure}[!htb]
	\centering
	\includegraphics[width=\linewidth]{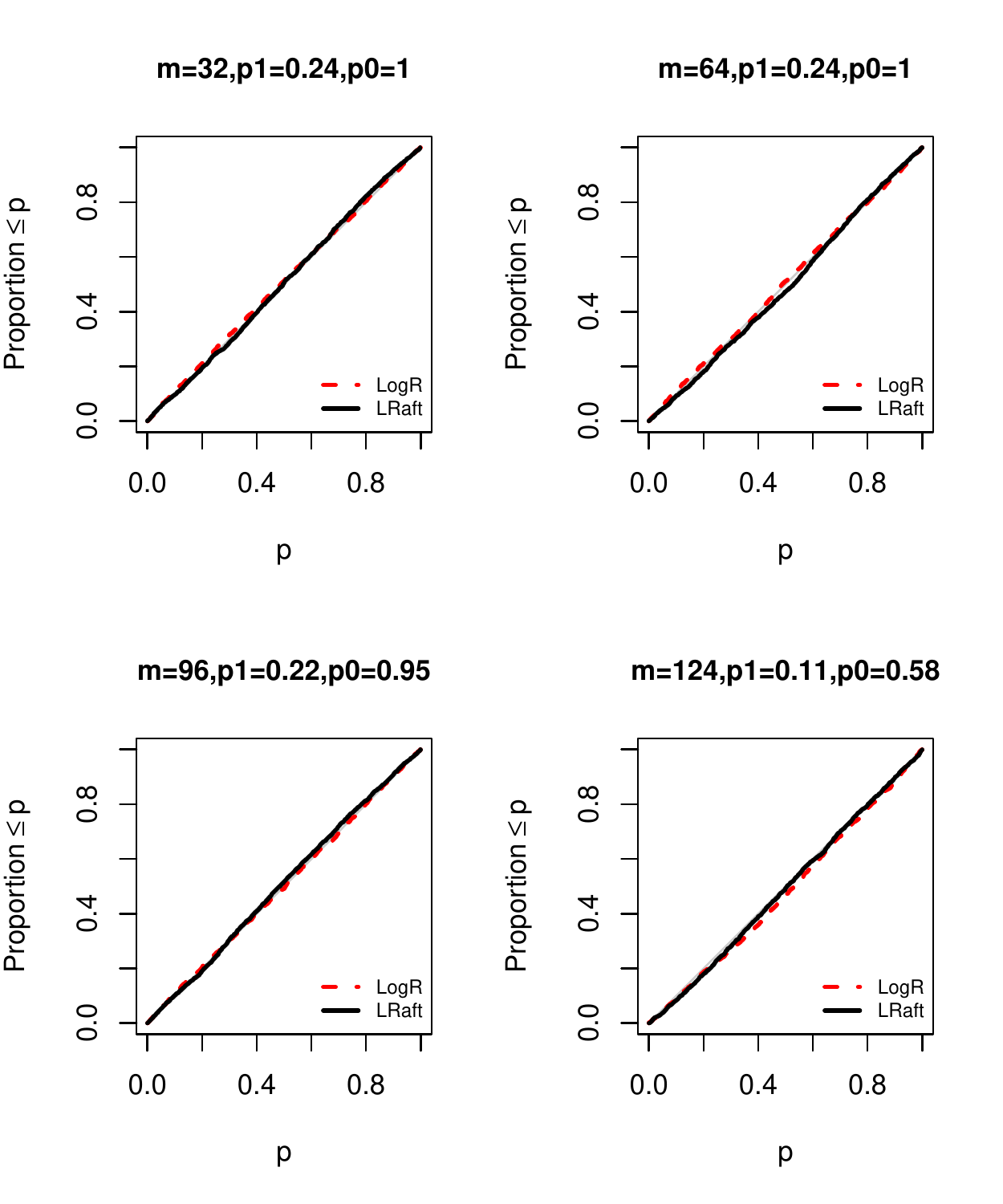}	
	\caption{Empirical cumulative distributions of p-values using the PA interference structure and $k=0.6$, with failure times correlated between individuals and with censoring times. Each panel corresponds to a different value of $m$ as stated in the title. The average proportions of observed failures in the $Z=1$ and $Z=0$ groups are stated in the title as `p1' and `p0' respectively.
	\label{sims-withcens-corr-add-PA-0}
	}
	\end{figure}

\begin{figure}[!htb]
	\centering		
	\includegraphics[width=\linewidth]{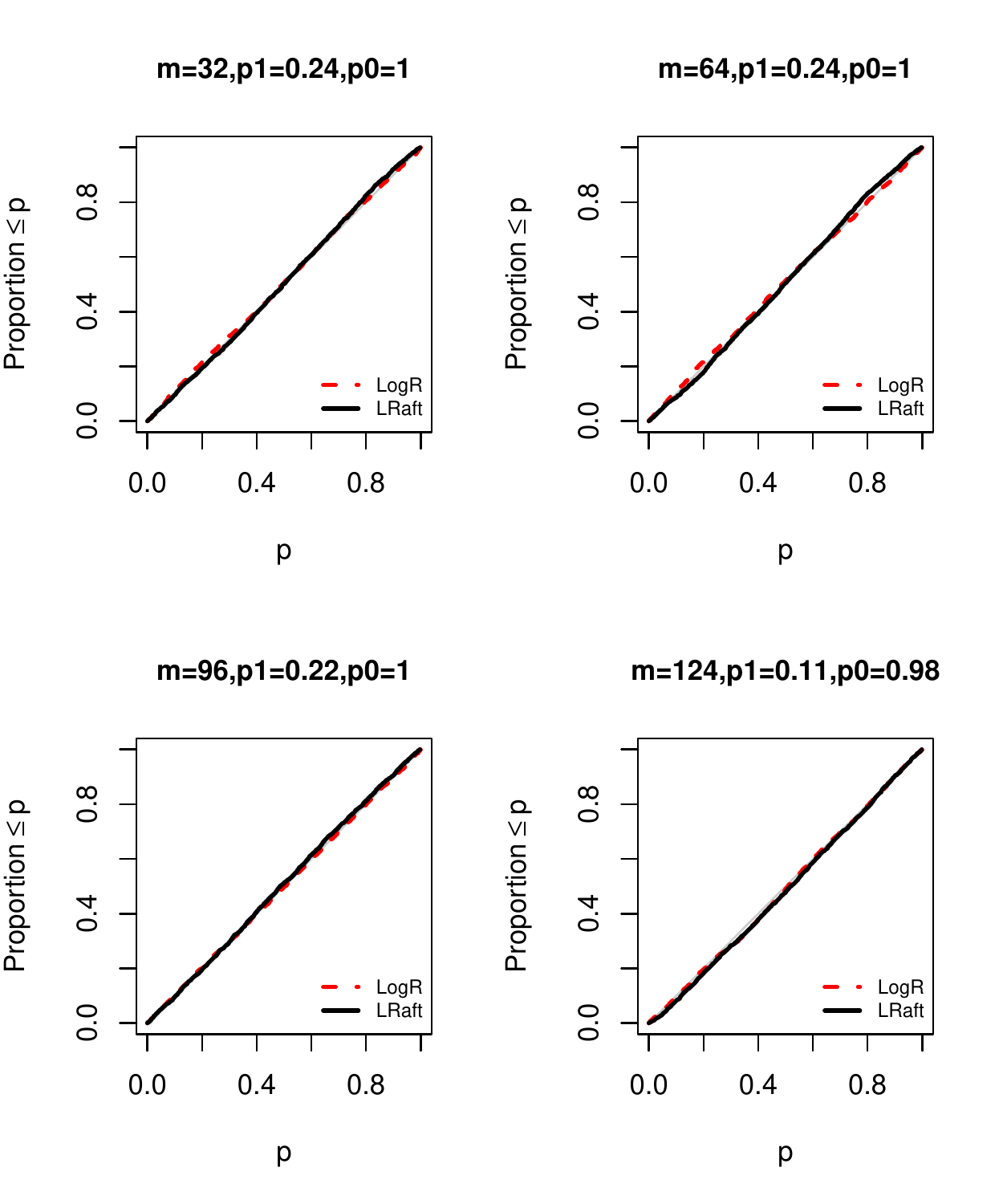}	
	\caption{Empirical cumulative distributions of p-values using the PA interference structure and $k=1$, with failure times correlated between individuals and with censoring times. Each panel corresponds to a different value of $m$ as stated in the title. The average proportions of observed failures in the $Z=1$ and $Z=0$ groups are stated in the title as `p1' and `p0' respectively.
	\label{sims-withcens-corr-add-PA-1}
	}	
	\end{figure}

The simulation study was further carried out with uniformity trial failure times that are correlated between individuals, and with the censoring times. The uniformity failure times $\tilde y_i({\bf 0})$ and the dropout times $\tilde C_i$ are generated in Step 0 and 1 respectively as follows:
\begin{enumerate}\setcounter{enumi}{-1}
\item Determine the $n \times n$ correlation matrix $\rho$ as follows. For $i,j=1,\ldots,n, i \neq j$, let $\tilde A_{ij} = A_{ij}/A_{i} \times U_{ij}$, where $U_{ij} \sim {\rm Uniform}(0.9,1)$. Set $\rho_{ii}=1, i=1,\ldots,n$, and $\rho_{ij}=\left(\tilde A_{ij}+\tilde A_{ji}\right), i,j=1,\ldots,n, i \neq j$. (The sum $\tilde A_{ij}+\tilde A_{ji}$ ensures symmetry in $\rho$, while the random jitter $U_{ij}$ ensures that $\rho$ is positive definite.) Sample the vector of $n$ log-transformed uniformity failure times, denoted by ${\bf u} = (u_1, \ldots, u_n)$, from the multivariate normal distribution with mean vector $\mu{\bf 1}_n$, where ${\bf 1}_n$ is the vector of $n$ ones, and covariance matrix $\sigma^2 \rho$; i.e., ${\bf u} \sim {\cal N}_n(\mu{\bf 1}_n, \sigma^2\rho)$. Let $\mu=4.5$ and $\sigma^2=0.25^2$. Set the uniformity failure time for individual $i$ as ${\tilde y}_i({\bf 0}) = \exp(u_i)$.

\item For an observed treatment assignment, sample the vector of $n$ log-transformed dropout times, denoted by ${\bf R} = (R_1, \ldots, R_n)$, from the multivariate normal distribution with mean vector $\mu{\bf 1}_n + \tau^{\dag} {\bf G}$, where ${\bf G} = (G_1, \ldots, G_n)$, and covariance matrix $\omega^2 \rho$; i.e., ${\bf R} \sim {\cal N}_n(\mu{\bf 1}_n + \tau^{\dag} {\bf G}, \omega^2\rho)$. Let $\tau^{\dag}=2.8$ and $\omega^2=1-0.25^2$.
Determine the dropout times as ${\tilde C}_i = \exp(R_i)$.
\end{enumerate}
All 16 combinations of $k=0.6,1$, $m=124,96,64,32$ and the interference structure in Section~\ref{sect:example_n256censoring_power} and the linear preferential attachment network are considered. Plots of the ECDFs of the p-values are shown in Figures~\ref{sims-withcens-corr-add-LH-0} to \ref{sims-withcens-corr-add-PA-1}. 
The failure rates were between $11\%$ and $24\%$ in the $Z=1$ arm, and between $58\%$ and $99\%$ in the $Z=0$ arm.


\section{Simulation study of statistical power \label{sect:example_n256censoring_power-plots}}

\begin{figure}[!htb]
	\centering
    \includegraphics[width=.32\linewidth]{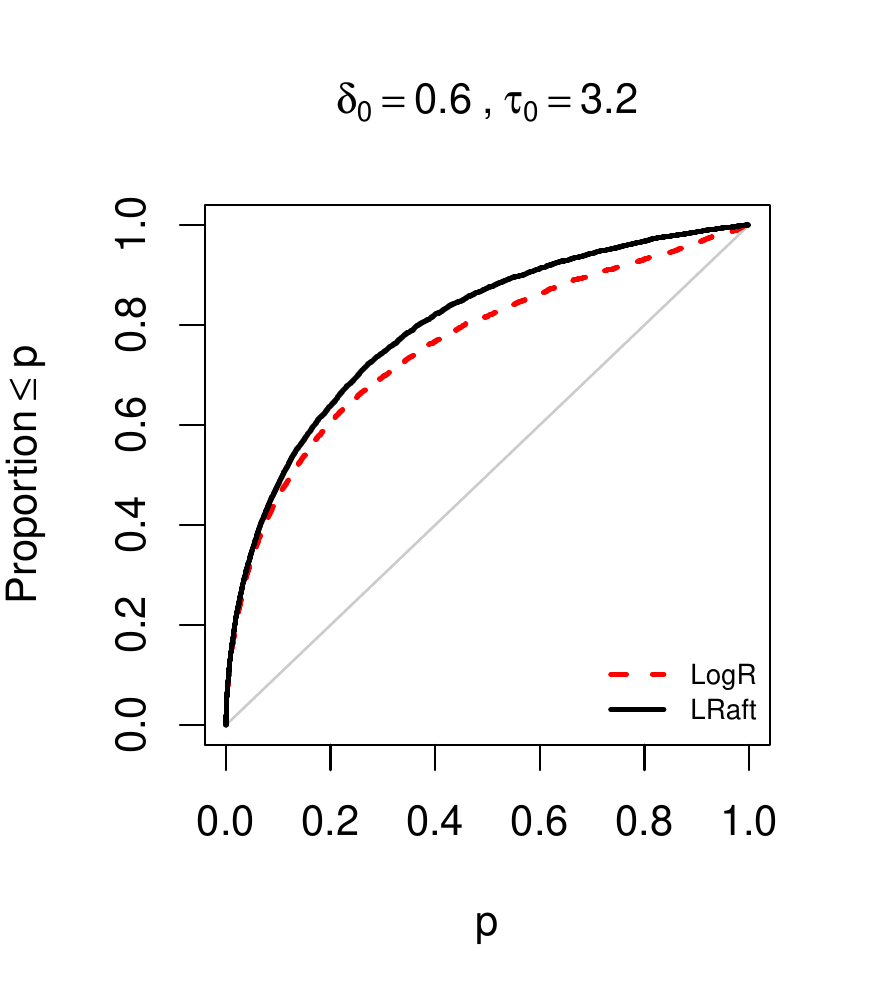}
    \includegraphics[width=.32\linewidth]{plot-power-n_256m96-8.pdf}    
    \includegraphics[width=.32\linewidth]{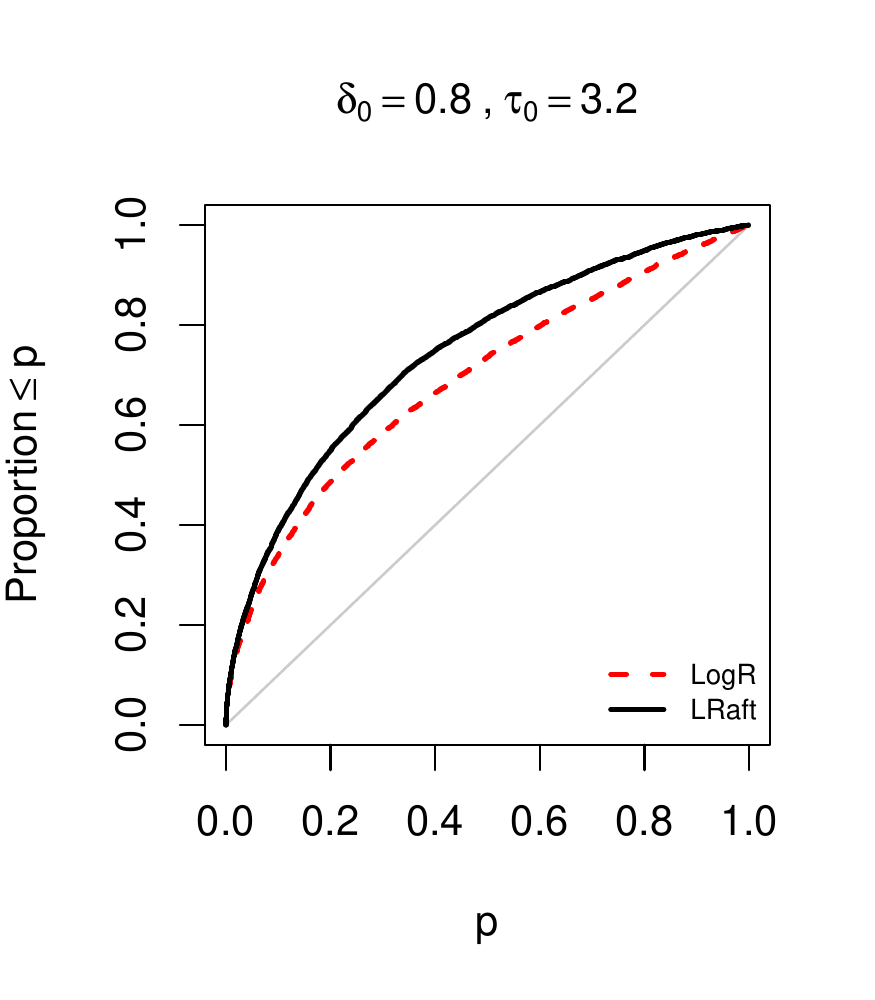}    
    \includegraphics[width=.32\linewidth]{plot-power-n_256m96-4.pdf}
    \includegraphics[width=.32\linewidth]{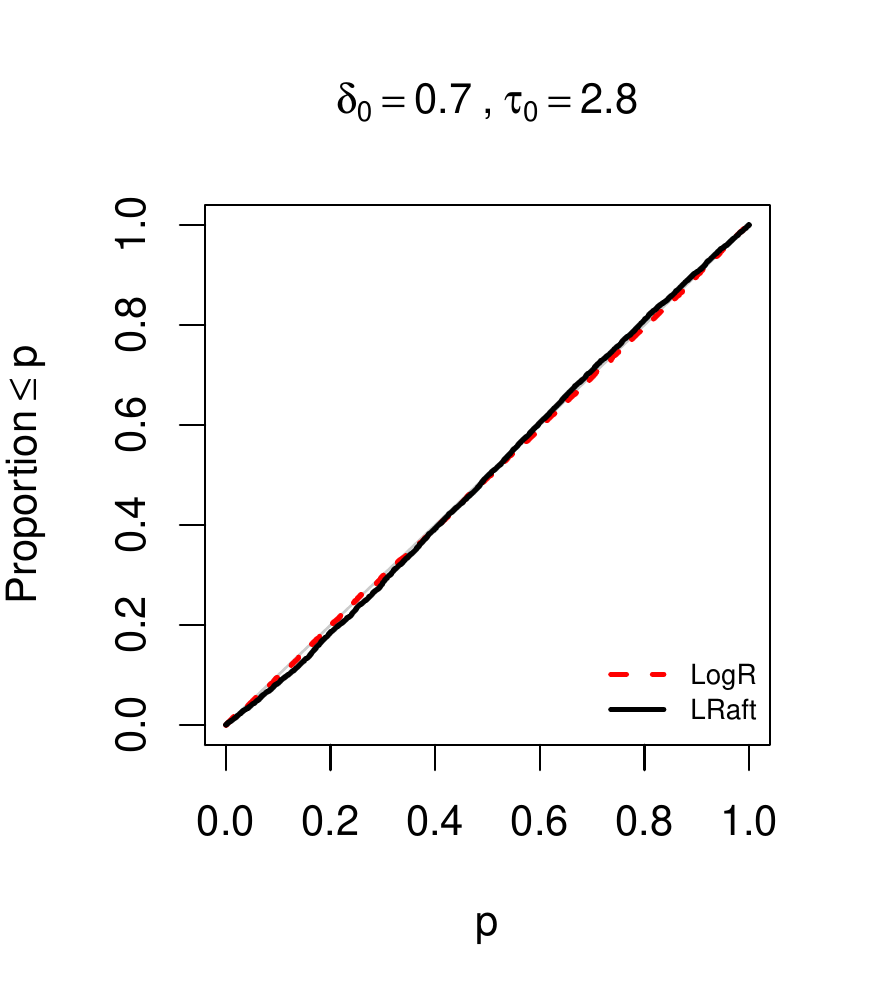}
    \includegraphics[width=.32\linewidth]{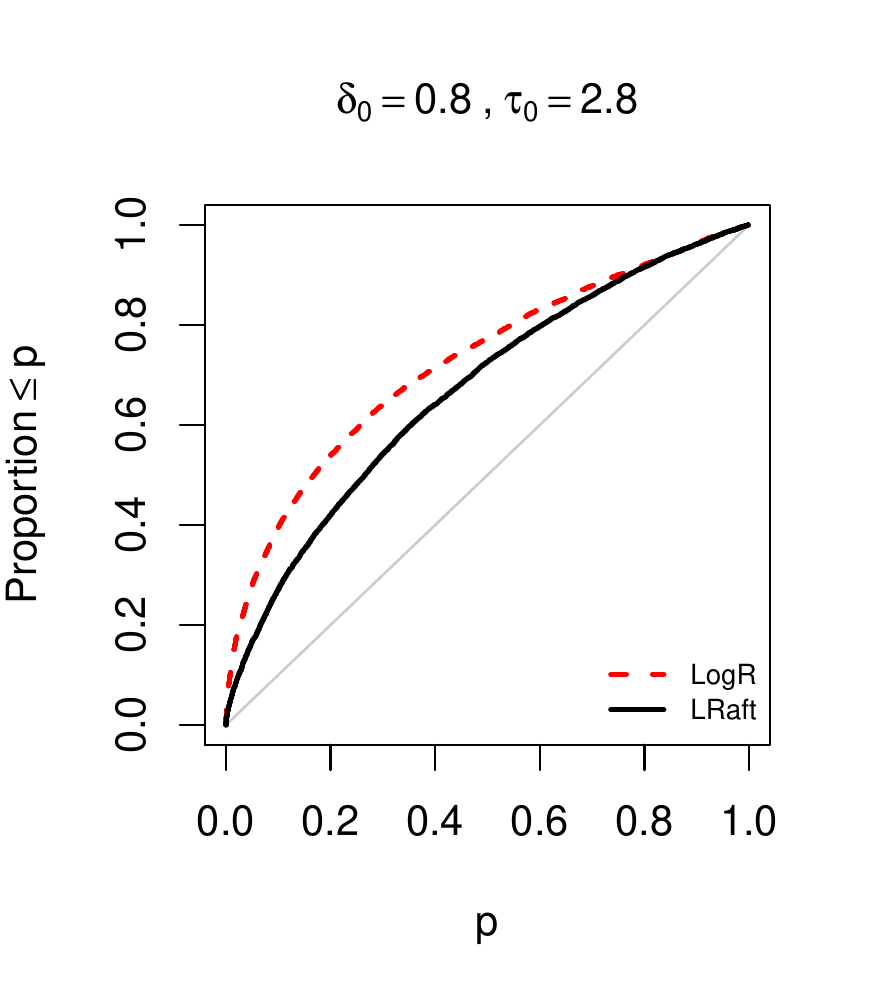}	
    \includegraphics[width=.32\linewidth]{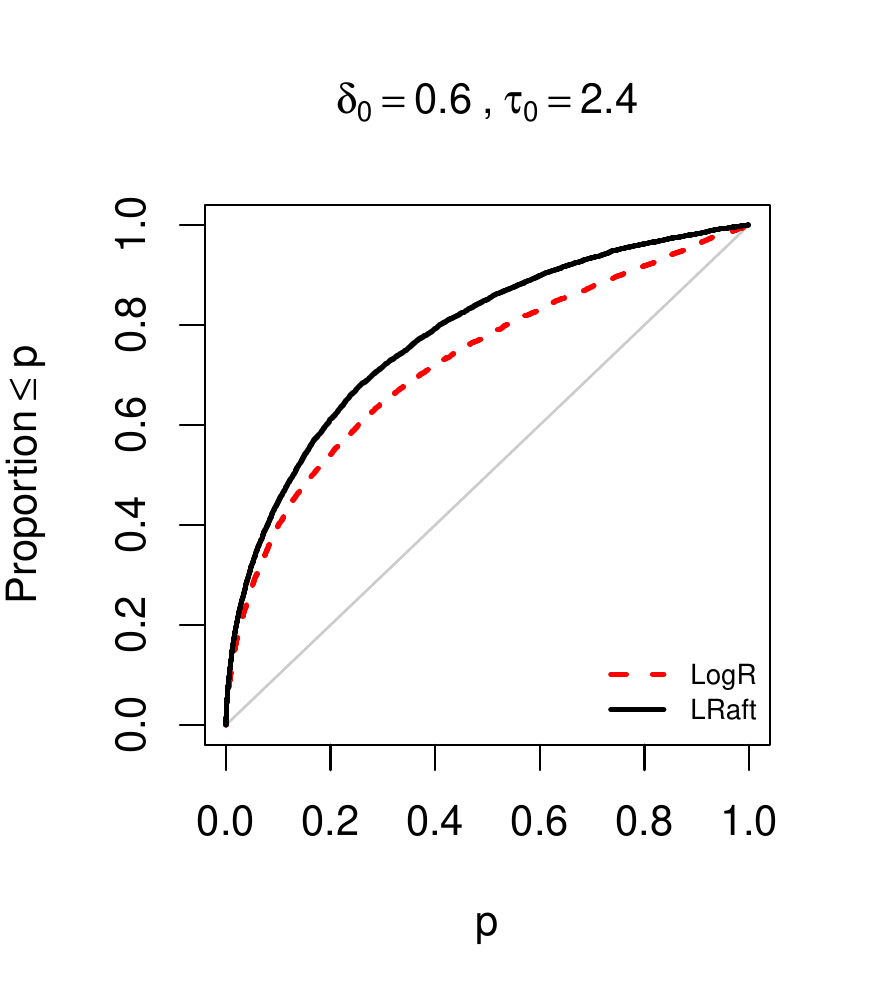}
    \includegraphics[width=.32\linewidth]{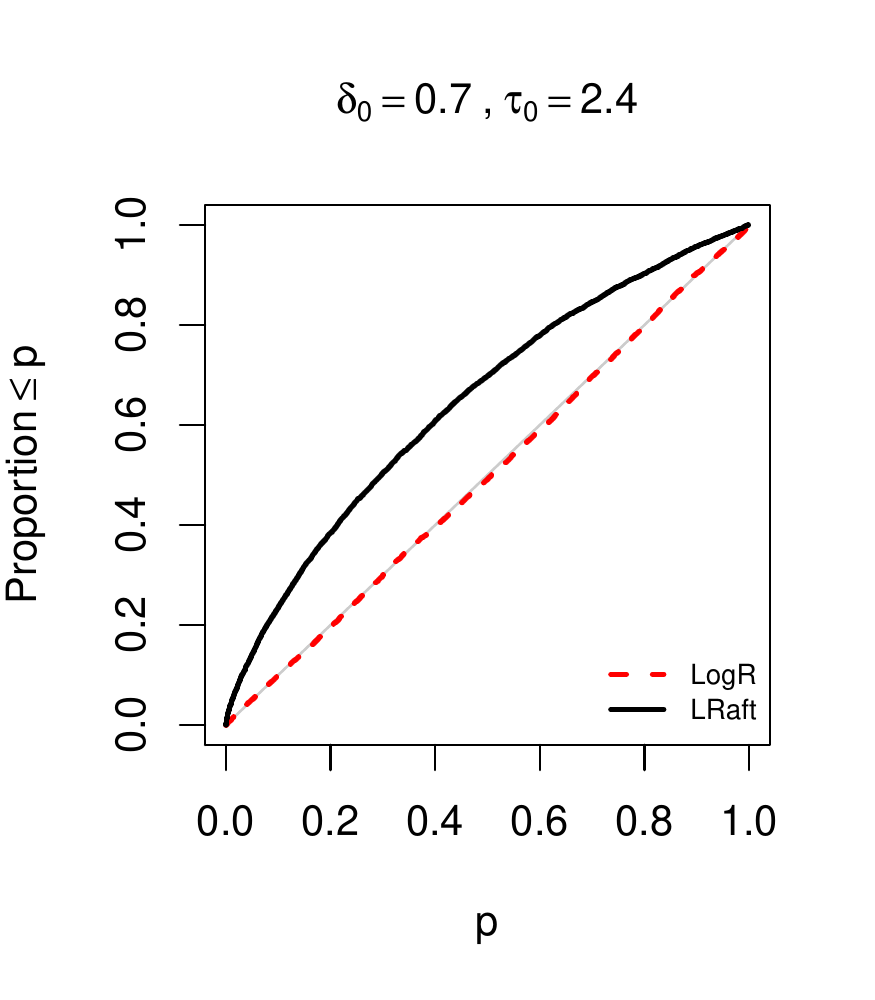}
    \includegraphics[width=.32\linewidth]{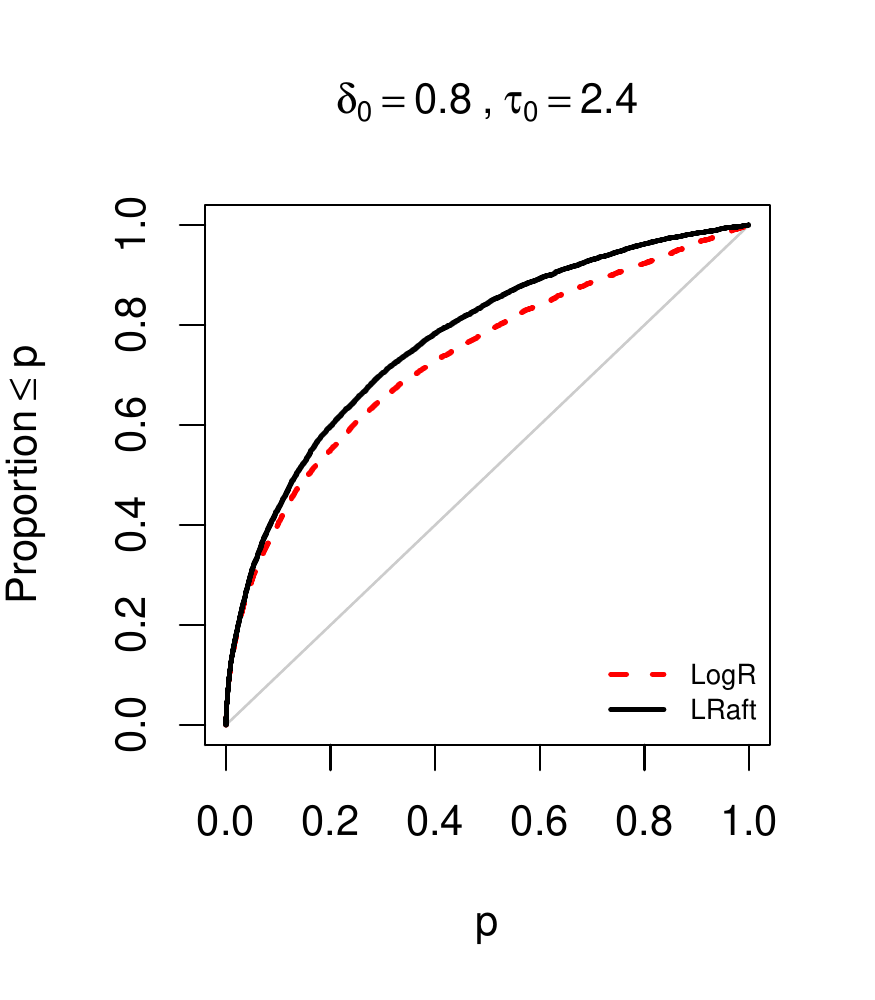}
	\caption{
	Empirical cumulative distributions of p-values from a simulation study using the proposed method in Section~\ref{sect:IPZ}. The true parameter values used to generate the data were $(\delta,\tau)=(0.7,2.8)$. The assumed values of $(\delta_0,\tau_0)$ for testing the null hypothesis are stated in the title of each panel. The panels corresponding to the null hypotheses $H_0: (\delta_0, \tau_0)=(0.6,2.8)$ and $H_0: (\delta_0, \tau_0)=(0.7,3.2)$ are presented in the main paper as the left and right panels respectively in 
Figure~\ref{plot-pvaluesECDF-complete_rand-sims-cens-n_128-m_64-d_07-t_28-}.	
	\label{plot-power-n_256m96}
	 }
	\end{figure}

\begin{figure}[!htb]
	\centering
    \includegraphics[width=\linewidth]
        {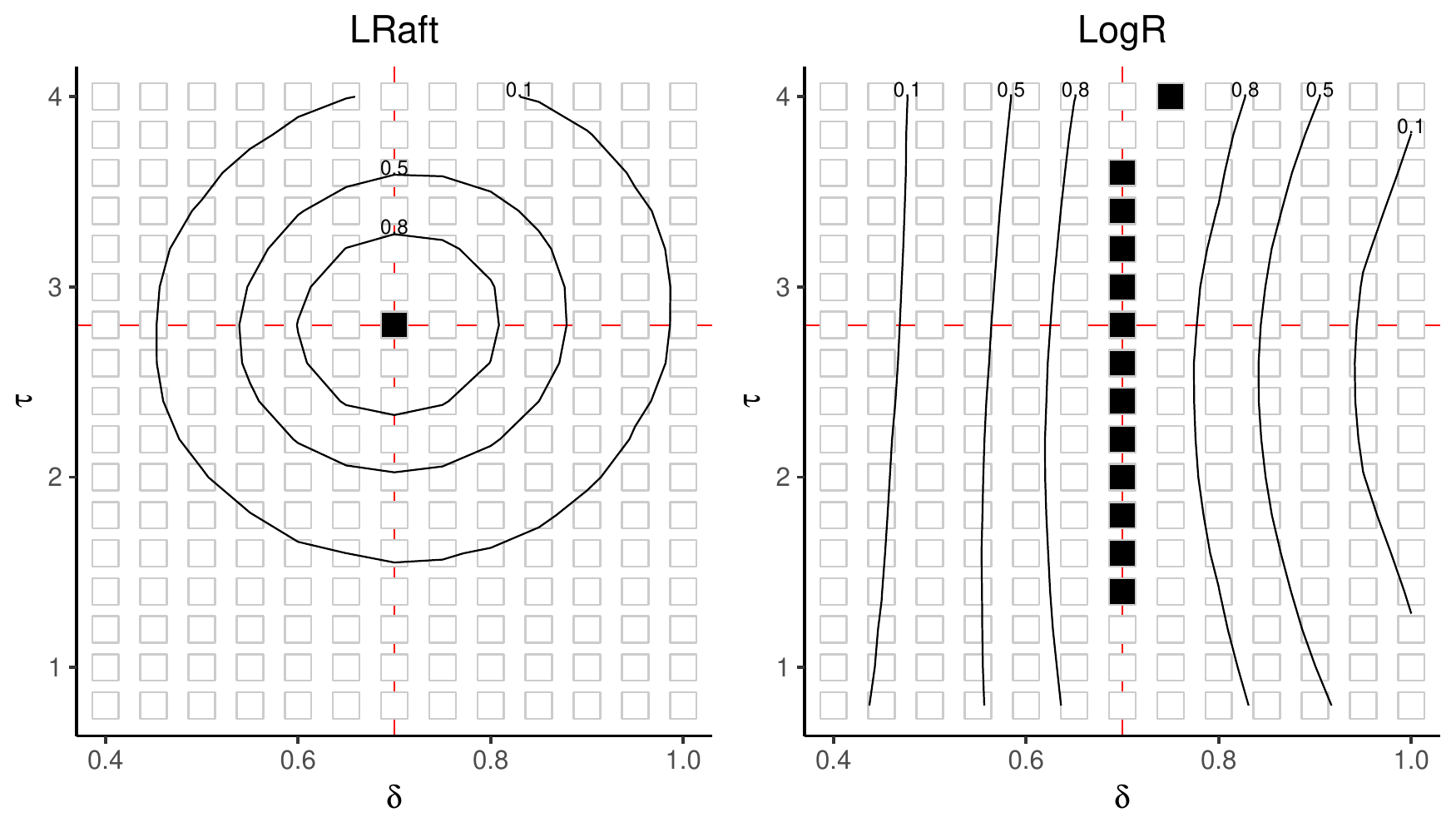}
	\caption{
	Average empirical coverage of 
	LRaft (left) and LogR (right) 95\% confidence sets.
	Each value of $(\delta_0, \tau_0)$ tested is indicated by a square,
	with the empirical coverage determined by the proportion of 95\% confidence sets 
	that included each pair of $(\delta_0, \tau_0)$.
	The contours are labelled with the coverage levels,
	with filled squares indicating coverage of at least 0.95. The true parameter values used to generate the data were $(\delta,\tau)=(0.7,2.8)$.
	\label{plot-complete_rand-sims-cens_TRUE-add_FALSE-model_2-n_256-modelH0_2-allsims-contours}
	}
	\end{figure}

To compare the power of the LRaft and LogR tests under the proposed method in Section~\ref{sect:IPZ}, the simulation study in Section~\ref{sect:example_n256censoring_power} was repeated as follows. For each simulated dataset with $k=1$ and $m=96$, the p-values ${\rm pv}^{\cal C}(\delta_{0},\tau_{0})$ with ${\cal C}=2500$ were calculated for a discrete grid of values $(\delta_0,\tau_0) \in \{0.4,0.5,\!\ldots\!,1.0\} \!\times\! \{ 0.8, 1.2 \!\ldots\!, 4.0\}$ in step 2.
Plots of the ECDFs of the p-values for selected values of $(\delta_0,\tau_0)$ are shown in Web Figure~\ref{plot-power-n_256m96}. 
95\% confidence sets for ($\delta,\tau$) were then constructed for each simulated dataset. The proportion of LRaft and LogR 95\% confidence sets that included each value of 
$(\delta_{0}, \tau_{0})$ tested are plotted in
Web Figure~\ref{plot-complete_rand-sims-cens_TRUE-add_FALSE-model_2-n_256-modelH0_2-allsims-contours}.
Both the LRaft and LogR confidence sets included the true value of $(\delta, \tau)=(0.7, 2.8)$ at the nominal coverage level. The LRaft confidence sets tended to exclude other values of $(\delta, \tau)$, except when $(\delta_0, \tau_0)$ were close to the true values $(\delta, \tau)=(0.7, 2.8)$. On the other hand, the LogR confidence sets tended to include, at the nominal coverage level, values of $(\delta_0, \tau_0)$ whenever $\delta_{0}$ was close to the true value of $\delta=0.7$, even for assumed values of $\tau_0$ which were not close to the true value of $\tau=2.8$. These results are in concert with those in Web Figure~\ref{plot-power-n_256m96} which show the LogR test lacks power to detect indirect (spillover) effects which are different from those specified under the null.

\section{Availability of \texttt{R} code}

The \texttt{R} code used to implement the proposed methods and to carry out the simulation studies in Section~\ref{sect:example_n256censoring_power} and in Web Appendices~\ref{sect:sims-withcens-add} and \ref{sect:example_n256censoring_power-plots} of this document are available at the following web address:
\url{
https://github.com/wwloh/General-Interference-Censoring
}

\end{document}